\documentclass[twocolumn]{aastex701}
\usepackage{graphicx}
\usepackage{amsmath}
\usepackage{hyperref}
\usepackage{longtable}
\usepackage{booktabs}

\newcommand{\pivec}{\mbox{\boldmath $\pi$}}
\newcommand{\muvec}{\mbox{\boldmath $\mu$}}

\newcommand{\te}{t_{\rm E}}
\newcommand{\thetae}{\theta_{\rm E}}

\newcommand{\pie}{\pi_{\rm E}}

\newcommand{\pien}{\pi_{{\rm E},N}}
\newcommand{\piee}{\pi_{{\rm E},E}}
\newcommand{\dl}{D_{\rm L}}

\def\e{{\rm E}}

\def\eqalign#1{\null\,\vcenter{\openup\jot
        \ialign{\strut\hfil$\displaystyle{##}$&$
        \displaystyle{{}##}$\hfil \crcr#1\crcr}}\,}
\definecolor{brown}{rgb}{0.59, 0.29, 0.0}
\definecolor{darkgreen}{rgb}{0.0, 0.42, 0.24}
\definecolor{darkblue}{rgb}{0.01, 0.31, 0.59}
\definecolor{blue}{rgb}{0.0,0.0,1.0}
\definecolor{green}{rgb}{0.0,1.0,0.0}

\begin{document}

\title{Candidate Microlensing Brown Dwarfs in Binary Lens Systems from the 2023--2025 Observing Seasons}
\shorttitle{Candidate Microlensing Brown Dwarfs}

% leading author =============================
\author{Cheongho Han}
\affiliation{Department of Physics, Chungbuk National University, Cheongju 28644, Republic of Korea}
\email{cheongho@astroph.chungbuk.ac.kr}
% -----------------
\author{Andrzej Udalski} 
\affiliation{Astronomical Observatory, University of Warsaw, Al.~Ujazdowskie 4, 00-478 Warszawa, Poland}
\email{udalski@astrouw.edu.pl} 
% -----------------
\author{Ian A. Bond}
\affiliation{School of Mathematical and Computational Sciences, Massey University, Auckland 0745, New Zealand}
\email{i.a.bond@massey.ac.nz}
% -----------------
\author{Chung-Uk Lee}
\affiliation{Korea Astronomy and Space Science Institute, Daejon 34055, Republic of Korea}
\email{leecu@kasi.re.kr}
\collaboration{14}{(Leading authors)}
% KMTNet ===========================
\author{Michael D. Albrow}   
\affiliation{University of Canterbury, Department of Physics and Astronomy, Private Bag 4800, Christchurch 8020, New Zealand}
\email{michael.albrow@canterbury.ac.nz}
% -----------------
\author{Sun-Ju Chung}
\affiliation{Korea Astronomy and Space Science Institute, Daejon 34055, Republic of Korea}
\email{sjchung@kasi.re.kr}
% -----------------
\author{Andrew Gould}
\affiliation{Department of Astronomy, Ohio State University, 140 West 18th Ave., Columbus, OH 43210, USA}
\email{gould.34@osu.edu}
% -----------------
\author{Youn Kil Jung}
\affiliation{Korea Astronomy and Space Science Institute, Daejon 34055, Republic of Korea}
\affiliation{University of Science and Technology, Daejeon 34113, Republic of Korea}
\email{younkil21@gmail.com}
% -----------------
\author{Kyu-Ha~Hwang}
\affiliation{Korea Astronomy and Space Science Institute, Daejon 34055, Republic of Korea}
\email{kyuha@kasi.re.kr}
% -----------------
\author{Yoon-Hyun Ryu}
\affiliation{Korea Astronomy and Space Science Institute, Daejon 34055, Republic of Korea}
\email{yhryu@kasi.re.kr}
% -----------------
\author{Yossi Shvartzvald}
\affiliation{Department of Particle Physics and Astrophysics, Weizmann Institute of Science, Rehovot 76100, Israel}
\email{yossishv@gmail.com}
% -----------------
\author{In-Gu Shin}
\affiliation{Department of Astronomy, Westlake University, Hangzhou 310030, Zhejiang Province, China}
\email{ingushin@gmail.com}
% -----------------
\author{Jennifer C. Yee}
\affiliation{Center for Astrophysics $|$ Harvard \& Smithsonian 60 Garden St., Cambridge, MA 02138, USA}
\email{jyee@cfa.harvard.edu}
% -----------------
\author{Weicheng Zang}
\affiliation{Department of Astronomy, Westlake University, Hangzhou 310030, Zhejiang Province, China}
\email{zangweicheng@westlake.edu.cn}
% -----------------
\author{Hongjing Yang}
\affiliation{Department of Astronomy, Westlake University, Hangzhou 310030, Zhejiang Province, China}
\email{yanghongjing@westlake.edu.cn}
% -----------------
\author{Doeon Kim}
\affiliation{Department of Physics, Chungbuk National University, Cheongju 28644, Republic of Korea}
\email{qso21@hanmail.net}
% -----------------
\author{Dong-Jin Kim}
\affiliation{Korea Astronomy and Space Science Institute, Daejon 34055, Republic of Korea}
\email{keaton03@kasi.re.kr}
% -----------------
\author{Seung-Lee Kim}
\affiliation{Korea Astronomy and Space Science Institute, Daejon 34055, Republic of Korea}
\email{slkim@kasi.re.kr}
% -----------------
\author{Dong-Joo Lee}
\affiliation{Korea Astronomy and Space Science Institute, Daejon 34055, Republic of Korea}
\email{marin678@kasi.re.kr}
% -----------------
\author{Sang-Mok Cha}
\affiliation{Korea Astronomy and Space Science Institute, Daejon 34055, Republic of Korea}
\email{chasm@kasi.re.kr}
% -----------------
\author{Yongseok Lee}
\affiliation{Korea Astronomy and Space Science Institute, Daejon 34055, Republic of Korea}
\email{yslee@kasi.re.kr}
% -----------------
\author{Byeong-Gon Park}
\affiliation{Korea Astronomy and Space Science Institute, Daejon 34055, Republic of Korea}
\email{bgpark@kasi.re.kr}
% -----------------
\author{Richard W. Pogge}
\affiliation{Department of Astronomy, Ohio State University, 140 West 18th Ave., Columbus, OH 43210, USA}
\email{pogge.1@osu.edu}
% -----------------
\collaboration{20}{(KMTNet Collaboration)}
%%%% OGLE ===========================
\author{Przemek Mr{\'o}z}
\affiliation{Astronomical Observatory, University of Warsaw, Al.~Ujazdowskie 4, 00-478 Warszawa, Poland}
\email{pmroz@astrouw.edu.pl}
% -----------------
\author{Micha{\l} K. Szyma{\'n}ski}
\affiliation{Astronomical Observatory, University of Warsaw, Al.~Ujazdowskie 4, 00-478 Warszawa, Poland}
\email{msz@astrouw.edu.pl}
% -----------------
\author{Jan Skowron}
\affiliation{Astronomical Observatory, University of Warsaw, Al.~Ujazdowskie 4, 00-478 Warszawa, Poland}
\email{jskowron@astrouw.edu.pl}
% -----------------
\author{Rados{\l}aw Poleski} 
\affiliation{Astronomical Observatory, University of Warsaw, Al.~Ujazdowskie 4, 00-478 Warszawa, Poland}
\email{radek.poleski@gmail.co}
% -----------------
\author{Igor Soszy{\'n}ski}
\affiliation{Astronomical Observatory, University of Warsaw, Al.~Ujazdowskie 4, 00-478 Warszawa, Poland}
\email{soszynsk@astrouw.edu.pl}
% -----------------
\author{Pawe{\l} Pietrukowicz}
\affiliation{Astronomical Observatory, University of Warsaw, Al.~Ujazdowskie 4, 00-478 Warszawa, Poland}
\email{pietruk@astrouw.edu.pl}
% -----------------
\author{Szymon Koz{\l}owski} 
\affiliation{Astronomical Observatory, University of Warsaw, Al.~Ujazdowskie 4, 00-478 Warszawa, Poland}
\email{simkoz@astrouw.edu.pl}
% -----------------
\author{Krzysztof A. Rybicki}
\affiliation{Astronomical Observatory, University of Warsaw, Al.~Ujazdowskie 4, 00-478 Warszawa, Poland}
\email{krybicki@astrouw.edu.pl}
% -----------------
\author{Patryk Iwanek}
\affiliation{Astronomical Observatory, University of Warsaw, Al.~Ujazdowskie 4, 00-478 Warszawa, Poland}
\email{piwanek@astrouw.edu.pl}
% -----------------
\author{Krzysztof Ulaczyk}
\affiliation{Department of Physics, University of Warwick, Gibbet Hill Road, Coventry, CV4 7AL, UK}
\email{kulaczyk@astrouw.edu.pl}
% -----------------
\author{Marcin Wrona}
\affiliation{Astronomical Observatory, University of Warsaw, Al.~Ujazdowskie 4, 00-478 Warszawa, Poland}
\affiliation{Villanova University, Department of Astrophysics and Planetary Sciences, 800 Lancaster Ave., Villanova, PA 19085, USA}
\email{mwrona@astrouw.edu.pl}
% -----------------
\author{Mariusz Gromadzki}          
\affiliation{Astronomical Observatory, University of Warsaw, Al.~Ujazdowskie 4, 00-478 Warszawa, Poland}
\email{marg@astrouw.edu.pl}
% -----------------
\author{Mateusz J. Mr{\'o}z} 
\affiliation{Astronomical Observatory, University of Warsaw, Al.~Ujazdowskie 4, 00-478 Warszawa, Poland}
\email{mmroz@astrouw.edu.pl}
\collaboration{100}{(The OGLE Collaboration)}
% ------------
%%%% MOA ===========================
\author{Fumio Abe}
\affiliation{Institute for Space-Earth Environmental Research, Nagoya University, Nagoya 464-8601, Japan}
\email{abe@isee.nagoya-u.ac.jp}
% -----------------
\author{David P. Bennett}
\affiliation{Code 667, NASA Goddard Space Flight Center, Greenbelt, MD 20771, USA}
\affiliation{Department of Astronomy, University of Maryland, College Park, MD 20742, USA}
\email{bennett.moa@gmail.com}
% -----------------
\author{Aparna Bhattacharya}
\affiliation{Code 667, NASA Goddard Space Flight Center, Greenbelt, MD 20771, USA}
\affiliation{Department of Astronomy, University of Maryland, College Park, MD 20742, USA}
\email{aparna.bhattacharya@nasa.gov}
% -----------------
\author{Ryusei Hamada}
\affiliation{Department of Earth and Space Science, Graduate School of Science, Osaka University, Toyonaka, Osaka 560-0043, Japan}
\email{hryusei@iral.ess.sci.osaka-u.ac.jp}
% -----------------
\author{Yuki Hirao}
\affiliation{Institute of Astronomy, Graduate School of Science, The University of Tokyo, 2-21-1 Osawa, Mitaka, Tokyo 181-0015, Japan}
\email{hirao@ioa.s.u-tokyo.ac.jp}
% -----------------
\author{Asahi Idei}
\affiliation{Department of Earth and Space Science, Graduate School of Science, Osaka University, Toyonaka, Osaka 560-0043, Japan}
\email{idei@iral.ess.sci.osaka-u.ac.jp}
% -----------------
\author{Stela Ishitani Silva}  
\affiliation{Code 667, NASA Goddard Space Flight Center, Greenbelt, MD 20771, USA}
\email{ishitanisilva@cua.edu}
% -----------------
\author{Shuma Makida}
\affiliation{Department of Earth and Space Science, Graduate School of Science, Osaka University, Toyonaka, Osaka 560-0043, Japan}
\email{makida@iral.ess.sci.osaka-u.ac.jp}
% -----------------
\author{Shota Miyazaki}
\affiliation{Institute of Space and Astronautical Science, Japan Aerospace Exploration Agency, 3-1-1 Yoshinodai, Chuo, Sagamihara, Kanagawa 252-5210, Japan}
\email{miyazaki@ir.isas.jaxa.jp}
% -----------------
\author{Yasushi Muraki}
\affiliation{Institute for Space-Earth Environmental Research, Nagoya University, Nagoya 464-8601, Japan}
\email{muraki@isee.nagoya-u.ac.jp}
% -----------------
\author{Tutumi Nagai}
\affiliation{Department of Earth and Space Science, Graduate School of Science, Osaka University, Toyonaka, Osaka 560-0043, Japan}
\email{nagai@iral.ess.sci.osaka-u.ac.jp}
% -----------------
\author{Togo Nagano}
\affiliation{Department of Earth and Space Science, Graduate School of Science, Osaka University, Toyonaka, Osaka 560-0043, Japan}
\email{nagano@iral.ess.sci.osaka-u.ac.jp}
% -----------------
\author{Seiya Nakayama}
\affiliation{Department of Earth and Space Science, Graduate School of Science, Osaka University, Toyonaka, Osaka 560-0043, Japan}
\email{nakayama@iral.ess.sci.osaka-u.ac.jp}
% -----------------
\author{Mayu Nishio}
\affiliation{Department of Earth and Space Science, Graduate School of Science, Osaka University, Toyonaka, Osaka 560-0043, Japan}
\email{nishio@iral.ess.sci.osaka-u.ac.jp}
% -----------------
\author{Kansuke Nunota}
\affiliation{Department of Earth and Space Science, Graduate School of Science, Osaka University, Toyonaka, Osaka 560-0043, Japan}
\email{unota@iral.ess.sci.osaka-u.ac.jp}
% -----------------
\author{Ryo Ogawa}
\affiliation{Department of Earth and Space Science, Graduate School of Science, Osaka University, Toyonaka, Osaka 560-0043, Japan}
\email{rogawa@iral.ess.sci.osaka-u.ac.jp}
% -----------------
\author{Ryunosuke Oishi}
\affiliation{Department of Earth and Space Science, Graduate School of Science, Osaka University, Toyonaka, Osaka 560-0043, Japan}
\email{oishi@iral.ess.sci.osaka-u.ac.jp}
% -----------------
\author{Yui Okumoto}
\affiliation{Department of Earth and Space Science, Graduate School of Science, Osaka University, Toyonaka, Osaka 560-0043, Japan}
\email{yuokumoto@iral.ess.sci.osaka-u.ac.jp}
% -----------------
\author{Greg Olmschenk}
\affiliation{Code 667, NASA Goddard Space Flight Center, Greenbelt, MD 20771, USA}
\email{greg@olmschenk.com}
% -----------------
\author{Cl{\'e}ment Ranc}
\affiliation{Sorbonne Universit\'e, CNRS, UMR 7095, Institut d'Astrophysique de Paris, 98 bis bd Arago, 75014 Paris, France}
\email{ranc@iap.fr}
% -----------------
\author{Nicholas J. Rattenbury}
\affiliation{Department of Physics, University of Auckland, Private Bag 92019, Auckland, New Zealand}
\email{n.rattenbury@auckland.ac.nz}
% -----------------
\author{Yuki Satoh}
\affiliation{College of Science and Engineering, Kanto Gakuin University, Yokohama, Kanagawa 236-8501, Japan}
\email{yukisato@kanto-gakuin.ac.jp}
% -----------------
\author{Takahiro Sumi}
\affiliation{Department of Earth and Space Science, Graduate School of Science, Osaka University, Toyonaka, Osaka 560-0043, Japan}
\email{sumi@ess.sci.osaka-u.ac.jp}
% -----------------
\author{Daisuke Suzuki}
\affiliation{Department of Earth and Space Science, Graduate School of Science, Osaka University, Toyonaka, Osaka 560-0043, Japan}
\email{dsuzuki@ir.isas.jaxa.jp}
% -----------------
\author{Takuto Tamaoki}
\affiliation{Department of Earth and Space Science, Graduate School of Science, Osaka University, Toyonaka, Osaka 560-0043, Japan}
\email{tamaoki@iral.ess.sci.osaka-u.ac.jp}
% -----------------
\author{Sean K. Terry}
\affiliation{Code 667, NASA Goddard Space Flight Center, Greenbelt, MD 20771, USA}
\affiliation{Department of Astronomy, University of Maryland, College Park, MD 20742, USA}
\email{skterry@umd.edu}
% -----------------
\author{Paul J. Tristram}
\affiliation{University of Canterbury Mt.~John Observatory, P.O. Box 56, Lake Tekapo 8770, New Zealand}
\email{tristram.p@gmail.com}
% -----------------
\author{Aikaterini Vandorou}
\affiliation{Code 667, NASA Goddard Space Flight Center, Greenbelt, MD 20771, USA}
\affiliation{Department of Astronomy, University of Maryland, College Park, MD 20742, USA}
\email{aikaterini.vandorou@utas.edu.au}
% -----------------
\author{Hibiki Yama}
\affiliation{Department of Earth and Space Science, Graduate School of Science, Osaka University, Toyonaka, Osaka 560-0043, Japan}
\email{yama@iral.ess.sci.osaka-u.ac.jp}
\collaboration{100}{(The MOA/PRIME Collaboration)}
% ------------
\correspondingauthor{\texttt{cheongho@astroph.chungbuk.ac.kr}}
\correspondingauthor{\texttt{leecu@kasi.re.kr}}

\begin{abstract}
We present detailed light-curve analyses of  ten binary-lens microlensing events observed 
during the 2023--2025 seasons and selected as candidates for hosting brown-dwarf companions. 
The sample includes OGLE-2023-BLG-0249, KMT-2023-BLG-1246, OGLE-2023-BLG-0079, KMT-2024-BLG-0072, 
KMT-2024-BLG-0897, KMT-2024-BLG-1876, KMT-2024-BLG-2379,  KMT-2025-BLG-0922, KMT-2025-BLG-1056, 
and KMT-2025-BLG-2427.  For each event, we carry out modeling of the light curve, explore 
relevant degeneracies, and, when finite-source effects are present, determine the angular 
Einstein radius. For OGLE-2023-BLG-0249, we additionally measure the microlens parallax, 
which allows a direct determination of the lens masses and distance. For the remaining 
events, we estimate the physical lens properties via Bayesian analyses incorporating 
Galactic priors. The resulting posteriors show that the lens companions in all systems 
have median masses in the brown-dwarf regime, and the lenses of two events (KMT-2025-BLG-0922 
and KMT-2025-BLG-1056) are consistent with binaries in which both lens components fall within 
the brown-dwarf mass range. Spanning a wide range of projected separations and distances, 
these detections illustrate the power of high-cadence microlensing surveys to build a census 
of brown-dwarf companions, including faint and distant systems beyond the reach of flux-limited 
methods. 
\end{abstract}


\section{Introduction \label{sec:one}} 

Brown dwarfs (BDs) occupy the mass range between stars and planets, making them key objects
for understanding where stellar formation ends and planetary formation begins \citep{Chabrier2000}. 
As substellar objects incapable of sustaining stable hydrogen fusion, BDs provide stringent
tests of theories for the formation and evolution of very low-mass objects, thereby refining 
our broader understanding of star formation. In addition, their cool, molecule-rich atmospheres
resemble those of giant exoplanets, allowing BDs to serve as accessible analogs for exoplanet
atmosphere studies and for calibrating atmospheric and chemical models \citep{Marley2015,
Kirkpatrick2005}. A comprehensive census of BDs is also essential for constraining the low-mass
end of the Galactic mass function and for assessing the contribution of substellar objects to the
mass budget of the Milky Way.

Microlensing provides a powerful and complementary approach for detecting BDs because it is
sensitive to mass rather than luminosity.  In contrast to direct imaging and spectroscopic 
techniques, microlensing does not require detection of light from the BD itself, enabling 
the discovery of objects that are intrinsically faint, cold, and/or distant. Consequently, 
microlensing is well suited for probing the Galactic population of low-mass objects and for 
constraining their occurrence rates and mass distribution.

\citet{Gould2022b} reported a systematic finite-source point-lens (FSPL) survey of giant-source 
microlensing events in the 2016--2019 data, yielding a homogeneous sample of 30 events. The 
authors found a pronounced gap in the angular Einstein radius ($\thetae$) distribution, $8.8 
<\thetae/\mu{\rm as} < 26$, which they call the ``Einstein Desert.'' Below this desert, four 
events cluster at very small $\thetae$, indicating a distinct population of low-mass free-floating 
planet candidates. Just above the desert, the events are dominated by stellar and BD lenses, 
producing a pile-up of BD/stars immediately above the Einstein Desert in the FSPL sample. This 
clear separation suggests that there are two different groups of lenses: BDs and stars 
produce events just above the Einstein Desert, while a separate population of much lower-mass 
free-floating planets produces the events below the desert.

Recent microlensing studies have also provided new insights into the so-called ``BD desert.'' 
Using a sample of binary-lens microlensing events, \citet{Zhang2025} performed a statistical 
analysis and identified a companion mass-ratio desert at $0.02 \lesssim q \lesssim 0.05$ for 
projected separations of $\sim$1--5~au.  Although the BD sample used in this analysis is not 
complete,\footnote{At present, the only homogeneous microlensing sample that probes the BD 
mass-ratio regime was presented by \citet{Shvartzvald2016} based on Wise, OGLE, and MOA data, 
although the number of events in that sample remains small.} the result indicates that the 
desert persists in microlensing-selected samples and further suggests that the mass ratio may 
provide a more robust and physically meaningful descriptor of the BD desert than the companion 
mass alone.

% Table 1 ------------------------------------------------------
\begin{deluxetable}{lllllll}
\tabletypesize{\footnotesize}
\tablewidth{0pt}
\tablecaption{Coordinates and event ID correspondence. \label{table:one}}
\tablehead{
\multicolumn{1}{c}{KMTNet ID}                 &
\multicolumn{1}{c}{(RA, DEC)$_{\rm J2000}$}   &
\multicolumn{1}{c}{$(l, b)$}                  &
\multicolumn{1}{c}{OGLE ID }                  &        
\multicolumn{1}{c}{MOA/PRIME  ID}                       
}
\startdata
 KMT-2023-BLG-0429        &  (18:16:05.27, -25:34:13.51)  &  ($+6^\circ$\hskip-2pt.2818, $-4^\circ$\hskip-2pt.2135)   &  {\bf OGLE-2023-BLG-0249}   &  MOA-2023-BLG-108     \\
 {\bf KMT-2023-BLG-1246}  &  (17:34:11.98, -27:40:04.40)  &  ($-0^\circ$\hskip-2pt.2608, $+2^\circ$\hskip-2pt.8034)   &  \nodata                    &  MOA-2023-BLG-271     \\
 KMT-2023-BLG-2750        &  (17:41:50.44, -25:41:58.96)  &  ($+2^\circ$\hskip-2pt.3172, $+2^\circ$\hskip-2pt.4110)   &  {\bf OGLE-2023-BLG-0079}   &  \nodata              \\
 {\bf KMT-2024-BLG-0072}  &  (17:39:02.33, -24:22:52.00)  &  ($+3^\circ$\hskip-2pt.1020, $+3^\circ$\hskip-2pt.6468)   &  \nodata                    &  \nodata              \\  
 {\bf KMT-2024-BLG-0897}  &  (17:51:46.90, -30:48:02.92)  &  ($-0^\circ$\hskip-2pt.9099, $-2^\circ$\hskip-2pt.1055)   &  \nodata                    &  \nodata              \\  
 {\bf KMT-2024-BLG-1876}  &  (17:53:20.99, -29:36:25.42)  &  ($+0^\circ$\hskip-2pt.2911, $-1^\circ$\hskip-2pt.7908)   &  OGLE-2024-BLG-0969         &  MOA-2024-BLG-163     \\  
 {\bf KMT-2024-BLG-2379}  &  (17:52:45.39, -29:42:57.49)  &  ($+0^\circ$\hskip-2pt.1317, $-1^\circ$\hskip-2pt.7350)   &  \nodata                    &  \nodata              \\  
 {\bf KMT-2025-BLG-0922}  &  (18:05:50.10, -32:16:32.92)  &  ($-0^\circ$\hskip-2pt.7065, $-5^\circ$\hskip-2pt.4370)   &  \nodata                    &  PRIME-2025-BLG-0156  \\  
 {\bf KMT-2025-BLG-1056}  &  (17:49:40.66, -29:40:24.89)  &  ($-0^\circ$\hskip-2pt.1734, $-1^\circ$\hskip-2pt.1385)   &  OGLE-2025-BLG-0764         &  \nodata              \\  
 {\bf KMT-2025-BLG-2427}  &  (17:32:46.65, -27:21:03.74)  &  ($-0^\circ$\hskip-2pt.1655, $+3^\circ$\hskip-2pt.2401)   &  OGLE-2025-BLG-1415         &  MOA-2025-BLG-0344   
\enddata                                                                                                                     
%\tablecomments{}
\end{deluxetable}
% --------------------------------------------------------------

In microlensing surveys, BD candidates can be identified through three primary observational
channels. The first channel consists of short-timescale single-lens events. Because the event
timescale scales as the square root of the lens mass, events lasting only a few days or less 
are preferentially produced by low-mass lenses such as BDs. However, short timescales can also 
arise from a large lens--source relative proper motion, and thus confirming a BD lens generally 
requires an additional constraint on the relative proper motion, e.g., \citet{Han2020}.

The second channel involves binary-lens events with very small mass ratios, e.g., \citet{Han2023, 
Han2024}.  Because most Galactic microlensing events are produced by low-mass stellar primaries 
\citep{Han2003}, a sufficiently small mass ratio strongly suggests that the secondary companion 
lies in the BD regime.

The third channel comprises binary-lens events that exhibit both short 
timescales and small angular Einstein radii, e.g., \citet{Han2025b}. While a short timescale 
alone suggests a low lens mass, a small $\thetae$, which likewise scales with the square root 
of the lens mass, provides an independent and stronger indication of a BD lens.  For binary-lens 
events, the likelihood of measuring $\thetae$ is relatively high because their light curves 
usually display caustic-related features that enable a determination of $\thetae$.  With $\thetae$ 
measured, the lens mass can be more tightly constrained, and if the microlens parallax is additionally 
measured, the lens mass can be uniquely determined, thereby enabling a definitive confirmation of 
the BD nature of the lens, e.g., \citet{Gould2009, Choi2013, Han2017b}.

In this work, we report candidate BDs in binaries identified from analyses of binary-lens events 
detected during the 2023--2025 seasons: OGLE-2023-BLG-0249, KMT-2023-BLG-1246, OGLE-2023-BLG-0079, 
KMT-2024-BLG-0072, KMT-2024-BLG-0897, KMT-2024-BLG-1876, KMT-2024-BLG-2379,  KMT-2025-BLG-0922, 
KMT-2025-BLG-1056, and KMT-2025-BLG-2427.  For each event, we carry out detailed light-curve 
modeling to determine the binary-lens parameters and to evaluate the probability that the 
companion lies in the BD regime, incorporating all available 
constraints.

The paper is organized as follows. 
Section~\ref{sec:two} describes the observations and data reduction. 
Section~\ref{sec:three} presents the modeling procedure and parameter estimation. 
Section~\ref{sec:four} characterizes the source stars and determines the angular Einstein radii. 
Section~\ref{sec:five} reports the inferred physical properties of the lens systems. 
Section~\ref{sec:six} discusses the feasibility of estimating the masses of the brown-dwarf 
candidates using future high-resolution observations.
Section~\ref{sec:seven} discusses the implications and summarizes our conclusions.

\section{Observations and data \label{sec:two}} 

Candidate BD companions in binaries were identified through systematic analyses of 
binary-lens microlensing events observed by the Korea Microlensing Telescope Network 
\citep[KMTNet;][]{Kim2016} survey during the 2023--2025 seasons. Table~\ref{table:one} 
summarizes the  ten events and lists their equatorial and Galactic coordinates.  Seven 
of these events were also monitored by other microlensing surveys, including the Optical 
Gravitational Lensing Experiment \citep[OGLE;][]{Udalski2015}, Microlensing Observations 
in Astrophysics \citep[MOA;][]{Bond2001, Sumi2003}, and PRime-focus Infrared Microlensing 
Experiment \citep[PRIME;][]{Sumi2025}, and the corresponding survey-specific event IDs 
are given in Table~\ref{table:one}.  For two events, OGLE-2023-BLG-0249 and OGLE-2023-BLG-0079, 
OGLE issued the initial alert prior to the KMTNet identification, and we therefore adopt the 
OGLE designations as the primary references for these events.  In Table~\ref{table:one}, 
primary references are indicated in boldface.

The events were selected through the following three channels. The first channel consists 
of binary-lens events with small mass ratios; in this work, we imposed an upper limit of 
$q_{\rm max} < 0.1$.  The second channel consists of short-timescale binary-lens events, 
independent of the mass ratio; we required the timescale to satisfy $t_{\rm E}\lesssim 10$~days.  
Such short timescales allow for the possibility that both lens components are BDs, implying 
a binary BD lens. The third channel includes events for which the lens mass could be uniquely 
determined and the inferred companion mass lies in the BD regime.

We emphasize that the present sample is not intended to be statistically complete.
The selection criteria were specifically designed to identify promising BD
candidates by focusing on events with observational characteristics that are sensitive
to low-mass companions, such as small mass ratios, short timescales, and measurable
finite-source effects. As a result, the sample is inherently biased toward systems in
which the companion is likely to lie in the BD mass regime. Consequently, the fact that
all selected events yield companion masses below the hydrogen-burning limit should not
be interpreted as representative of the underlying population, but rather as a natural
outcome of the targeted selection strategy. A broader exploration of parameter space,
including events with larger mass ratios or longer timescales, would likely reveal a
more diverse population spanning stellar, substellar, and planetary companions, although
with increased ambiguity in mass classification when strong constraints such as
$\theta_{\rm E}$ or $\pi_{\rm E}$ are not available.

The data used in our analyses were obtained from observations carried out by the four major 
microlensing surveys: KMTNet, OGLE-IV, MOA, and PRIME.  KMTNet operates three identical 
1.6-m wide-field telescopes located at Cerro Tololo Inter-American Observatory in Chile 
(KMTC), the South African Astronomical Observatory in South Africa (KMTS), and Siding 
Spring Observatory in Australia (KMTA), providing near-continuous longitudinal coverage; 
each is equipped with a mosaic camera yielding a field of view of $\sim$4~deg$^2$ for 
efficient high-cadence monitoring of dense Galactic bulge fields. OGLE-IV observations 
are carried out with the 1.3-m Warsaw Telescope at Las Campanas Observatory in Chile, 
using a wide-field mosaic CCD camera covering $\sim$1.4~deg$^2$. MOA observations are 
conducted with the 1.8-m MOA-II telescope at Mount John University Observatory in New 
Zealand, equipped with a wide-field CCD mosaic camera providing a field of view of $\sim$ 
2.2~deg$^2$.  Finally, PRIME utilizes the 1.8-m near-infrared telescope at the South African 
Astronomical Observatory equipped with a camera delivering a field of view of 1.45~deg$^2$.

KMTNet and OGLE observe primarily in the Cousins $I$ band, with additional $V$-band 
observations used to measure source colors. MOA typically observes in a custom broad 
red filter (MOA-$R$, roughly corresponding to Cousins $R+I$), supplemented by $V$-band 
data to obtain color information. PRIME observations were obtained primarily in the $H$ 
band.

The photometric error bars reported by the reduction pipeline are rescaled so that they 
properly represent the scatter of the data. We normalize the uncertainties by requiring 
that the reduced $\chi^2$ of the best-fit model for each data set is close to unity. The 
rescaled errors are then written as $\sigma^\prime = k\sqrt{\sigma^2 + \sigma_{\rm min}^2}$, 
where $k$ is a multiplicative scaling factor and $\sigma_{\rm min}$ represents an additional 
error floor that accounts for residual systematics. This rescaling prevents any single data 
set from being over- or under-weighted in the modeling and yields more reliable parameter 
estimates. The values of $k$ and $\sigma_{\rm min}$ are determined following the procedure 
of \citet{Yee2012}.

\section{Analyses \label{sec:three}} 

Because the light curves of the analyzed lensing events exhibit anomalous features associated 
with caustics, we model each event using a binary-lens single-source (2L1S) configuration. In 
this framework, a 2L1S light curve is described by seven basic parameters: $t_0$, $u_0$, $\te$, 
$s$, $q$, $\alpha$, and $\rho$. The first two parameters specify the lens--source geometry: 
$t_0$ is the time of closest approach, and $u_0$ is the projected lens–source separation at 
that time, scaled to $\thetae$. The parameter $\te$ is the event timescale, defined as the 
time required for the source to traverse an Einstein radius. The binary-lens properties are 
characterized by $s$ and $q$ (binary parameters), the projected separation and mass ratio of 
the two lens components, respectively. The parameter $\alpha$ denotes the angle between the 
source trajectory and the binary-lens axis. Finally, $\rho \equiv \theta_*/\thetae$, the ratio 
of the angular source radius to the angular Einstein radius, is required to model the magnification 
during caustic crossings or close approaches to caustics.

For some events, the standard model based on the seven basic parameters leaves long-term
residuals. In such cases, we fit extended models that include higher-order effects. We 
consider two effects: (1) the microlens-parallax effect \citep{Gould1992} and (2) lens-orbital 
motion \citep{Dominik1998, Albrow2000, Skowron2011}. The microlens-parallax effect arises from 
Earth's orbital motion around the Sun, whereas the lens-orbital effect arises from the orbital 
motion of the binary lens itself. To account for microlens parallax, we introduce two additional 
parameters $(\pien,\piee)$, which are the north and east components of the microlens-parallax 
vector, $\pivec_{\rm E}$. This vector is defined as $\pivec_{\rm E} = (\pi_{\rm rel}/ \thetae)
(\muvec/\mu)$, where $\pi_{\rm rel}=\pi_{\rm L}-\pi_{\rm S}$ is the relative lens--source parallax, 
and $\muvec$ (with magnitude $\mu$) is the relative lens--source proper-motion vector. For 
lens-orbital motion, under the approximation that the changes in the relative positions of the 
lens components are small over the duration of the event, we parameterize the orbital effects 
with two first-order terms: $ds/dt$ and $d\alpha/dt$. These represent the annual change rates 
of the binary separation and source trajectory angle, respectively.

In our modeling, we also explore potential degeneracies among solutions. To identify local 
minima, we perform a dense grid search over the binary-lens parameter space. When multiple 
solutions yield comparably good fits to the data, we report these alternatives and examine 
the origin of the degeneracy.

% Table 2 ------------------------------------------------------
\begin{deluxetable*}{lllllll}
%\tabletypesize{\scriptsize}
\tablewidth{0pt}
\tablecaption{Lensing parameters of OGLE-2023-BLG-0249 and KMT-2023-BLG-1246. \label{table:two}}
\tablehead{
\multicolumn{1}{c}{Parameter}           &
\multicolumn{2}{c}{OGLE-2023-BLG-0249}  &
\multicolumn{2}{c}{KMT-2023-BLG-1246 }  \\
\multicolumn{1}{c}{ }                   &
\multicolumn{1}{c}{$u_0 > 0$ }          &
\multicolumn{1}{c}{$u_0 < 0$ }          &
\multicolumn{1}{c}{Inner }              &
\multicolumn{1}{c}{Outer }   
}
\startdata
 $\chi^2$               &  $917.7               $  &  $918.3               $   &  $1576.9             $   &  $1576.0             $ \\
 $t_0$ (HJD$^\prime$)   &  $110.479 \pm 0.050   $  &  $110.329 \pm 0.046   $   &  $112.2621 \pm 0.0067$   &  $112.2311 \pm 0.0053$ \\
 $u_0$                  &  $0.7620 \pm 0.0008   $  &  $-0.7597 \pm 0.0013  $   &  $0.0173 \pm 0.0018  $   &  $0.0195 \pm 0.0024  $ \\
 $\te$ (days)           &  $52.00 \pm 0.17      $  &  $51.34 \pm 0.20      $   &  $22.27 \pm 1.73     $   &  $21.96 \pm 2.17     $ \\
 $s$                    &  $1.6993 \pm 0.0013   $  &  $1.6994 \pm 0.0017   $   &  $1.1973 \pm 0.013   $   &  $0.881 \pm 0.010    $ \\
 $q$                    &  $0.1299 \pm 0.0013   $  &  $0.132 \pm 0.0016    $   &  $0.01918 \pm 0.0018 $   &  $0.0155 \pm 0.0014  $ \\
 $\alpha$ (rad)         &  $1.90453 \pm 0.00057 $  &  $-1.90497 \pm 0.00081$   &  $4.909 \pm 0.010    $   &  $5.034 \pm 0.019    $ \\
 $\rho$ ($10^{-3}$)     &  $4.726 \pm 0.027     $  &  $4.789 \pm 0.029     $   &  $1.91 \pm 0.26      $   &  $1.73 \pm 0.21      $ \\
 $\pien$                &  $-0.280 \pm 0.028    $  &  $0.389 \pm 0.038     $   &   \nodata                &   \nodata              \\
 $\piee$                &  $-0.329 \pm 0.042    $  &  $-0.398 \pm 0.049    $   &   \nodata                &   \nodata              \\
 $ds/dt$                &  $-0.448 \pm 0.020    $  &  $-0.478 \pm 0.026    $   &   \nodata                &   \nodata              \\
 $d\alpha/dt$           &  $0.277 \pm 0.035     $  &  $-0.336 \pm 0.039    $   &   \nodata                &   \nodata              \\
\enddata
\tablecomments{
 ${\rm HJD}^\prime\equiv {\rm HJD}-2460000$.
}
\end{deluxetable*}
% --------------------------------------------------------------

\subsection{OGLE-2023-BLG-0249 \label{sec:three-one}}

The microlensing event OGLE-2023-BLG-0249 was first identified by the OGLE collaboration
during its early rising phase on 2021 April 7, corresponding to the reduced heliocentric
Julian date ${{\rm HJD}^\prime \equiv {\rm HJD}-2460000 = 41}$. It was later
independently detected by the KMTNet survey on April 17 (${\rm HJD}^\prime = 52$). The
source lies in the KMTNet field BLG31, which was monitored at a 2.5-hour cadence. The
$I$-band baseline magnitude of the event is $I_{\rm base}=15.58$, and the line-of-sight
extinction toward the field is $A_I=0.99$.

\begin{figure}[t]
\includegraphics[width=\columnwidth]{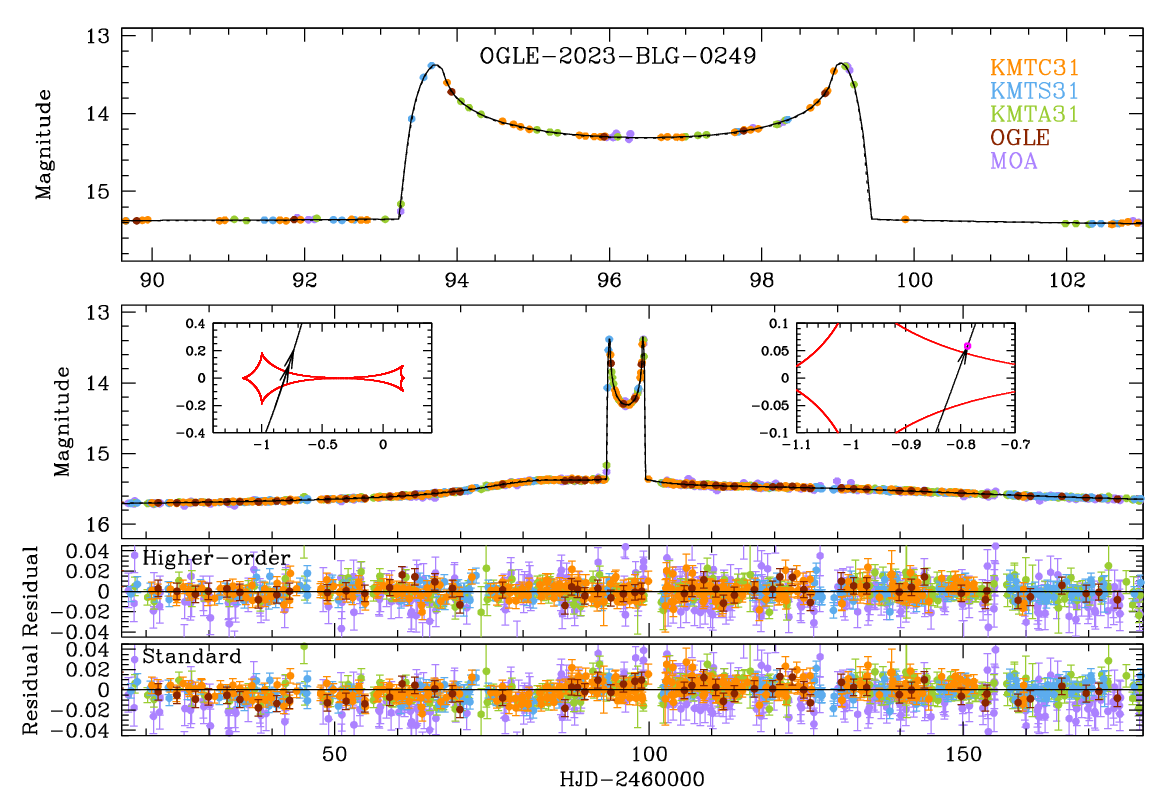}
\centering
\caption{
Light curve of OGLE-2023-BLG-0249. The upper two panels present the full light curve and 
a zoomed-in view focusing on caustic-related features. The lower two panels display the 
residuals relative to models without higher-order effects (standard model) and with 
higher-order effects included (higher-order model). Data points are color-coded according 
to the telescopes indicated in the legend. The insets in the second panel from the top 
illustrate the lens-system geometry, showing the source trajectory relative to the caustic 
structure. The left inset provides the overall configuration, while the right inset zooms 
in on the source's caustic-crossing regions.  The coordinates are centered on the binary-lens 
barycenter, with the abscissa aligned with the binary axis, and the more massive component 
lies to the right.
}
\label{fig:one}
\end{figure}

Figure~\ref{fig:one} displays the light curve of the OGLE-2023-BLG-0249. It shows the
characteristic morphology of a 2L1S event, including caustic-crossing features consisting 
of two sharp spikes at ${\rm HJD}^\prime \simeq 93.7$ and $\simeq 99.0$, with a 
characteristic U-shaped trough between them.  Both caustic spikes are well resolved by 
the combined data from multiple observatories. In addition, the light curve exhibits a
low-amplitude bump centered at ${\rm HJD}^\prime \simeq 80$.

% Figure 1 ------------------------------------------------------
% --------------------------------------------------------------

We initially model the light curve with a standard 2L1S framework, not considering 
higher-order effects. This yields a binary-lens solution with $(s,q)\sim (1.68, 0.10)$ 
and an event timescale of $\te\sim 56$~days. While this standard model provides an 
acceptable fit, it leaves coherent residuals persisting over much of the event, as shown 
in the residual panel of Figure~\ref{fig:one}.

Motivated by these long-lived residuals and the relatively long event duration, we next 
incorporate higher-order effects due to lens orbital motion and the Earth's orbital motion. 
This higher-order model substantially improves the fit, yielding $\Delta\chi^2=370.8$ 
relative to the standard model.  The corresponding model curve and residuals are shown in 
Figure~\ref{fig:one}.  We report two solutions with $u_0>0$ and $u_0<0$, arising from the 
ecliptic degeneracy \citep{Poindexter2005}.  The complete sets of lensing parameters for 
the two solutions are listed in Table~\ref{table:two}.  The two solutions are nearly 
indistinguishable, with the $u_0>0$ solution favored by only $\Delta\chi^2=0.6$. The 
inferred microlens parallax is $\pi_{\rm E}=(\pi_{{\rm E},E}^2+ \pi_{{\rm E},N}^2)^{1/2}
\sim 0.43$ for the $u_0>0$ solution and $\pi_{\rm E}\sim 0.56$ for the $u_0<0$ solution. 
The normalized source radius is tightly constrained by the well-resolved caustic crossings, 
$\rho\sim 4.7\times 10^{-3}$, enabling an estimate of the angular Einstein radius. As 
discussed in Sect.~\ref{sec:five}, the lens mass is uniquely determined, and the inferred 
companion lens mass places it in the BD regime.

With the lens mass ($M$) and distance ($D_{\rm L}$), together with the measured orbital 
parameters $(ds/dt, d\alpha/dt)$, we evaluate the ratio of projected kinetic to potential 
energy using the relation of \citet{Dong2009},
\begin{equation}
\left( {{\rm KE}\over {\rm PE}}\right)_\perp = 
{(a_\perp/{\rm au})^3 \over 8\pi^2(M/M_\odot) }
\left[ 
\left( {1\over s} {ds/dt \over {\rm yr}^{-1}}\right)^2 + 
\left( {d\alpha/dt \over {\rm yr}^{-1}}\right)^2
\right],
\label{eq1}
\end{equation}
\hskip-4pt
where $a_\perp = s D_{\rm L}\theta_{\rm E}$ is the projected physical separation. With 
the lens mass determined in Sect.~\ref{sec:five}, we find $({\rm KE}/{\rm PE})_\perp = 
0.15$ for the $u_0>0$ solution and $({\rm KE}/{\rm PE})_\perp = 0.14$ for the $u_0<0$ 
solution. For a bound system, this ratio should be less than unity, and both solutions 
satisfy this criterion.

The insets in the second panel from the top of Figure~\ref{fig:one} show the lens-system 
geometry for the $u_0>0$ solution. 
The geometry for the $u_0<0$ solution is nearly the mirror image of the $u_0>0$ case with 
respect to the binary axis. The binary lens produces two caustics, a larger caustic located 
near the lower-mass component ($M_2$) and a smaller caustic near the higher-mass component 
($M_1$), connected by a narrow bridge. The source crosses the caustic adjacent to $M_2$, 
generating the caustic spike at ${\rm HJD}^\prime \simeq 93.7$ upon entry and the spike at 
${\rm HJD}^\prime \simeq 99.0$ upon exit. The low-amplitude bump centered at 
${\rm HJD}^\prime \simeq 80$ arises from the source's approach to the lower off-axis cusp.

\subsection{KMT-2023-BLG-1246 \label{sec:three-two}} 

The microlensing event KMT-2023-BLG-1246 was first identified by the KMTNet survey on 2023 
June 14 (${\rm HJD}^\prime = 109$). Two days later, it was independently discovered by the 
MOA survey. The source lies in the KMTNet field BLG15, which was monitored with a one-hour 
cadence.

\begin{figure}[t]
\includegraphics[width=\columnwidth]{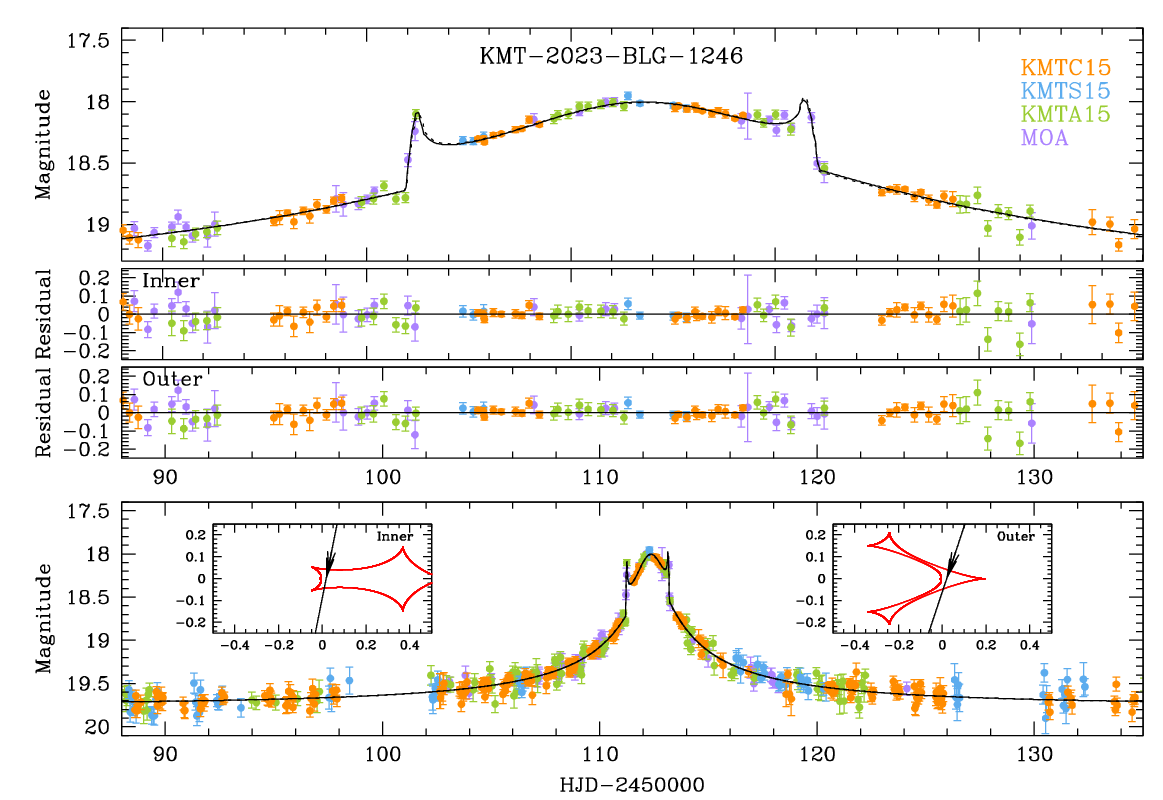}
\centering
\caption{
Light curve of KMT-2023-BLG-1246. The two insets in the bottom panel show the lens-system 
configuration of the inner and outer solutions. 
}
\label{fig:two}
\end{figure}

The light curve of the event is shown in Figure~\ref{fig:two}. The anomaly resembles those 
of the planetary events MOA-2022-BLG-091Lb and KMT-2024-BLG-1209 \citep{Han2025a} in that 
the caustic-crossing spikes are weak and the region between the spikes does not exhibit a 
characteristic U-shaped trough.  Instead, it approximately follows the magnification pattern 
expected from the underlying 1L1S event. Consequently, the anomaly was not readily recognizable 
in the online data and was confirmed after rereduction. The two caustic spikes at 
${\rm HJD}^\prime \simeq 102$ and 120 are partially resolved by the combined MOA and KMTA 
data.

Light-curve modeling indicates that the anomaly is well described by a 2L1S interpretation 
with a very small mass ratio, making the lens companion a BD candidate. We find two local 
solutions associated with the inner--outer degeneracy: $(s, q)_{\rm in}\sim (1.20, 0.019)$ 
and $(s, q)_{\rm out}\sim (0.88, 0.016)$, both with a similar event timescale of 
$\te\sim 22$~days. The full set of lensing parameters for the two solutions is listed in 
Table~\ref{table:two}. The two models are highly degenerate, with the outer solution favored 
by only $\Delta\chi^2=0.9$.  In Figure~\ref{fig:two}, we plot the model light curve for the 
outer solution over the data.

% Figure 2 ------------------------------------------------------
% --------------------------------------------------------------

% Table 3 ------------------------------------------------------
\begin{deluxetable*}{lllllll}
%\tabletypesize{\scriptsize}
\tablewidth{0pt}
\tablecaption{Lensing parameters of OGLE-2023-BLG-0079, KMT-2024-BLG-0072, and KMT-2024-BLG-0897. \label{table:three}}
\tablehead{
\multicolumn{1}{c}{Parameter}           &
\multicolumn{2}{c}{OGLE-2023-BLG-0079}  &
\multicolumn{1}{c}{KMT-2024-BLG-0072 }  &
\multicolumn{1}{c}{KMT-2024-BLG-0897 }  \\
\multicolumn{1}{c}{      }              &
\multicolumn{1}{c}{Inner }              &
\multicolumn{1}{c}{Outer }              &
\multicolumn{1}{c}{      }              &
\multicolumn{1}{c}{      }   
}
\startdata
 $\chi^2$               &  $2957.1           $   &  $2958.3           $   &  $295.9             $    &  $5998.4         $ \\
 $t_0$ (HJD$^\prime$)   &  $9999.51 \pm 0.26 $   &  $10000.14 \pm 0.26$   &  $381.273 \pm 0.066 $    &  $432.70 \pm 0.20$ \\
 $u_0$                  &  $0.239 \pm 0.015  $   &  $0.226 \pm 0.014  $   &  $-0.0046 \pm 0.0083$    &  $0.334 \pm 0.044$ \\
 $\te$ (days)           &  $25.43 \pm 0.85   $   &  $26.77 \pm 0.85   $   &  $7.34 \pm 0.14     $    &  $10.71 \pm 0.55 $ \\
 $s$                    &  $0.747 \pm 0.013  $   &  $1.129 \pm 0.020  $   &  $1.551 \pm 0.020   $    &  $0.961 \pm 0.012$ \\
 $q$                    &  $0.0282 \pm 0.0066$   &  $0.0375 \pm 0.0060$   &  $0.603 \pm 0.079   $    &  $0.094 \pm 0.025$ \\
 $\alpha$ (rad)         &  $2.335 \pm 0.081  $   &  $2.500 \pm 0.073  $   &  $5.964 \pm 0.026   $    &  $0.322 \pm 0.051$ \\
 $\rho$ ($10^{-3}$)     &  \nodata               &  \nodata               &  $< 14$                  &  $3.80 \pm 0.49  $ \\
\enddata
\tablecomments{
${\rm HJD}^\prime\equiv {\rm HJD}-2450000$ for OGLE-2023-BLG-0079 and
${\rm HJD}^\prime\equiv {\rm HJD}-2460000$ for the other events.
}
\end{deluxetable*}
% --------------------------------------------------------------

Within the context of this work, KMT-2023-BLG-1246 is a “BD candidate”, but it will also 
likely be included in KMTNet planet-host mass-ratio studies \citep{Zang2025}, which adopt 
the selection criterion $q\leq 0.03$. Whether the companion should be physically classified 
as a brown dwarf or a planet can be determined only by measuring the host mass, for example 
through late-time adaptive optics (AO) imaging \citep[e.g.,][]{Batista2015}.

The inner-outer degeneracy was originally identified for cases in which the source trajectory 
passes on the inner versus outer side of a peripheral caustic \citep{Gaudi1997}, and it was 
later extended to central perturbations by \citet{Yee2021} and \citet{Zhang2022}. \citet{Hwang2022} 
and \citet{Gould2022a} further showed that the two solutions satisfy an analytic relation in which 
the geometric mean of the separations is set by the lens--source separation at the epoch
of the anomaly,
\begin{equation}
\sqrt{s_{\rm in} \times s_{\rm out}} = s_{\pm}^{\dagger},\ \ \
s_{\pm}^{\dagger} = {\sqrt{u_{\rm anom}^2+4}\pm u_{\rm anom} \over 2}.
\label{eq2}
\end{equation}
\hskip-4pt
Here $s_{\rm in}$ and $s_{\rm out}$ are the binary separations of the inner and outer solutions,
respectively.  The quantity $u_{\rm anom}$ is the source--lens separation at the anomaly time 
($t_{\rm anom}$), i.e., 
\begin{equation}
u_{\rm anom} = \sqrt{u_0^2 + \tau^2_{\rm anom}}; \qquad
\tau_{\rm anom} = {t_{\rm anom}-t_0 \over \te}.
\label{eq3}
\end{equation}
\hskip-4pt
The  ``$+$'' branch in Eq.~(\ref{eq2}) applies to bump-like anomalies, while the ``$-$'' branch 
applies to dip-like anomalies. For KMT-2023-BLG-1246, we obtain $s^\dagger \sim 1.05$, consistent 
with the geometric mean of the fitted separations, $\sqrt{s_{\rm in}\times s_{\rm out}} \sim 1.03$.

The lens-system geometries corresponding to the inner and outer solutions are shown in the two
insets of the bottom panel of Figure~\ref{fig:two}. In both cases, the caustic forms a single 
resonant structure in which the central and planetary caustics merge. For the inner solution, 
the source trajectory passes on the inner side of the peripheral caustic, whereas for the outer 
solution it passes on the outer side.

% Figure 3 ------------------------------------------------------
\begin{figure}[t]
\includegraphics[width=\columnwidth]{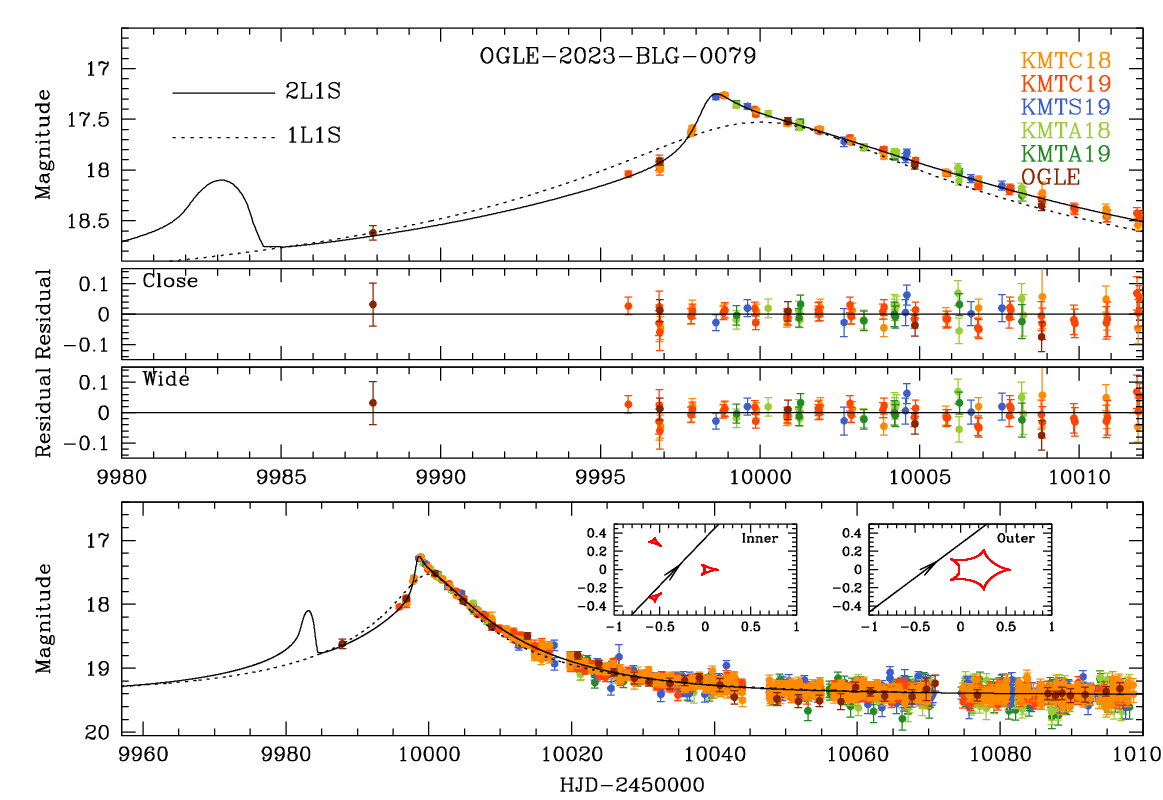}
\centering
\caption{
Light curve of OGLE-2023-BLG-0079. The notations are the same as in Fig.~\ref{fig:two}. The 
two insets in the bottom panel shows the lens-system configurations of the inner and outer
solutions.
}
\label{fig:three}
\end{figure}
% --------------------------------------------------------------

\subsection{OGLE-2023-BLG-0079 \label{sec:three-three}} 

The microlensing event OGLE-2023-BLG-0079 was first identified by the OGLE survey on 2023 
April 1 (${\rm HJD}^\prime \equiv {\rm HJD}-2450000= 10035$).  The event was subsequently 
recovered by KMTNet in in KMTNet's annual postseason analysis of the data \citep{Kim2018}.  
Because the magnification began before the start of the 2023 observing season, the resulting 
light curve is only partially covered. The source, with a baseline $I$-band magnitude of 
$I_{\rm base}=19.55$, lies in the overlap region of the KMTNet fields BLG18 and BLG19, each 
monitored with a cadence of $\sim 1$ hr.  The extinction toward the field is $A_I=2.51$.

Figure~\ref{fig:three} shows the light curve of the event.  Despite the incomplete coverage, 
it exhibits a clear asymmetry about the peak at ${\rm HJD}^\prime \simeq 9998.5$. The data 
near the peak show positive deviations, whereas the pre-peak portion of the light curve 
(${\rm HJD}^\prime \lesssim 9957$) exhibits negative deviations.

Light-curve modeling with a 2L1S configuration yields two degenerate solutions resulting 
from the inner--outer degeneracy.  In both solutions, the mass ratio is small, with $q \sim 
0.028$ for the inner solution and $q \sim 0.034$ for the outer solution, indicating that 
the lens companion is almost certainly either a BD or a planet.  Note, however, that 
because one solution has $q>0.03$, it is unlikely to to enter KMTNet planet-host mass-ratio 
samples \citep{Zang2025}.  The event timescale is $\te \sim 26$~days, and the normalized 
source radius could not be measured. The full set of best-fit lensing parameters for both 
the inner and outer solutions is listed in Table~\ref{table:three}. The two solutions are 
highly degenerate, with the inner solution favored by only $\Delta\chi^2 = 1.1$. The model 
light curve corresponding to the inner solution is shown in Figure~\ref{fig:three}.

The two insets in the bottom panel of Figure~\ref{fig:three} show the lens-system geometries 
for the inner and outer solutions. In the inner solution, the binary lens produces three 
caustics: one central caustic and two peripheral caustics. In contrast, the outer solution 
yields a single resonant caustic, in which the central and peripheral caustics merge. Although 
the caustic morphologies differ, the source trajectory probes the back-side region of the central 
caustic in both solutions. The positive deviation near the peak is produced as the source passes 
close to the upper cusp of the central caustic, whereas the negative deviation in the pre-peak 
phase arises from the source traversing the demagnification region located behind the central 
caustic.

\subsection{KMT-2024-BLG-0072 \label{sec:three-four}} 

The microlensing event KMT-2024-BLG-0072 was discovered on 2024 March 13 (${\rm HJD}^\prime 
\equiv {\rm HJD}-2460000 = 382$) and was observed exclusively by the KMTNet survey. The source, 
located in the KMTNet field BLG16, was monitored with a cadence of 2.5 hours.

\begin{figure}[t]
\includegraphics[width=\columnwidth]{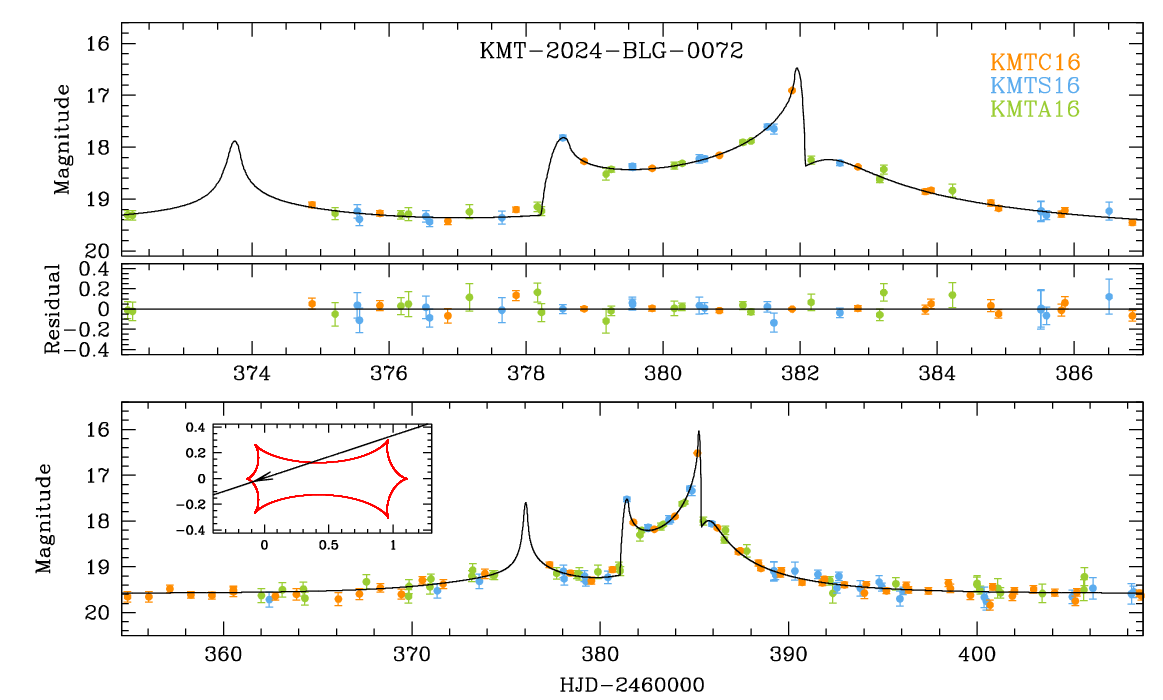}
\centering
\caption{
Lensing light curve of KMT-2024-BLG-0072.
}
\label{fig:four}
\end{figure}

The event light curve is shown in Figure~\ref{fig:four}. It exhibits two caustic-related 
features: a pair of caustic spikes at ${\rm HJD}^\prime \sim 378.5$ and 381.9, and a broader 
bump centered at ${\rm HJD}^\prime \sim 374$.  Owing to the relatively low observational 
cadence, neither of the caustic spikes was temporally resolved.

Modeling the light curve with a 2L1S configuration yields a unique solution that successfully
reproduces all of the anomalous features. The complete set of best-fit lensing parameters for 
this solution is listed in Table~\ref{table:three}. The mass ratio between the lens components 
($q \simeq 0.61$) is not very small. Nevertheless, we classify this event as a BD candidate 
because of its short event timescale, $\te \sim 7.3$ days. The best-fit model light curve is 
overplotted on the data in Figure~\ref{fig:four}.  Although the normalized source radius could 
not be determined accurately due to the relatively sparse coverage of the caustic crossings, 
an upper limit of $\rho_{\rm max} \sim 14 \times 10^{-3}$ could be constrained.

% Figure 4 ------------------------------------------------------
% --------------------------------------------------------------

The lens geometry is shown in the inset of the bottom panel of Figure~\ref{fig:four}. The binary 
lens produces a six-fold resonant caustic elongated along the binary axis. The source traverses 
the caustic diagonally, first passing near the upper-right off-axis cusp and producing the bump 
at ${\rm HJD}^\prime \sim 374$. It then enters and exits the caustic by crossing the upper-left 
and lower-left fold caustics, generating the two observed spikes. After exiting the caustic, the 
source approaches the left on-axis cusp, resulting in a weak bump approximately half a day later.

\subsection{KMT-2024-BLG-0897 \label{sec:three-five}} 

The lensing event KMT-2024-BLG-0897 was discovered on 2024 May 07 (${\rm HJD}^\prime \sim 437$) 
and was observed exclusively by the KMTNet survey. The event was short-lived, with the main 
magnification episode completed within 5 days. Despite its brief duration, the event was well 
covered because the source lay in the overlap region of the KMTNet prime fields BLG01 and BLG41, 
which were monitored at a combined cadence of 15 minutes.

% Figure 5 ------------------------------------------------------
\begin{figure}[t]
\includegraphics[width=\columnwidth]{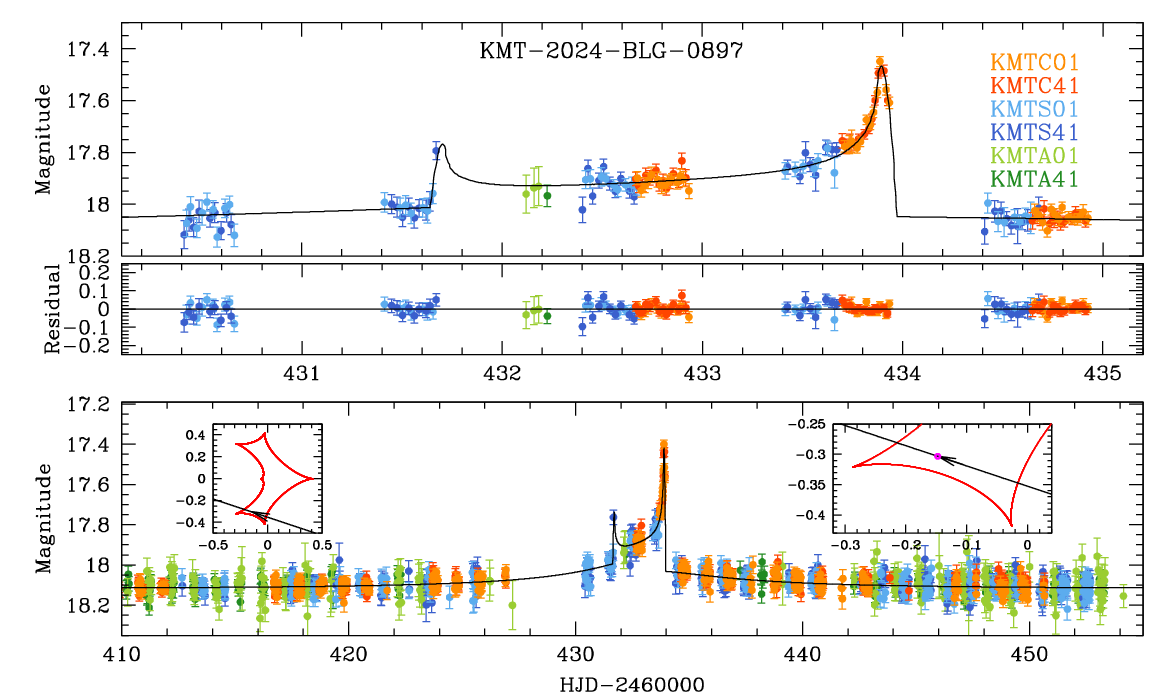}
\centering
\caption{
Lensing light curve of KMT-2024-BLG-0897. The insets in the bottom panel show the lens-system 
configuration. The left inset presents the full caustic structure, while the right inset 
provides a zoomed-in view of the region around the source's caustic crossings.
}
\label{fig:five}
\end{figure}
% --------------------------------------------------------------

The light curve of KMT-2024-BLG-0897 is shown in Figure~\ref{fig:five}. It exhibits prominent
caustic-crossing features, with two caustic spikes separated by $\sim 2.3$~days. The first 
spike was partially covered by the KMTS data, while the second spike was well resolved by the 
KMTC data.

Modeling the light curve yields a unique solution with binary-lens parameters $(s, q) \sim (0.96,
0.09)$ and an event timescale $\te \sim 10.7$~days. The small mass ratio, together with the short
timescale, makes the lens companion a BD candidate. Table~\ref{table:three} lists the full set of 
lensing parameters. The normalized source radius, $\rho \sim 3.8 \times 10^{-3}$, was measured 
from the resolved caustic-crossing features. The corresponding best-fit model is shown in 
Figure~\ref{fig:five} as a solid curve.

The lens-system configuration corresponding to the best-fit solution is shown in the insets 
of the bottom panel of Figure~\ref{fig:five}. The left inset illustrates the full caustic 
geometry, while the right inset provides an enlarged view of the region around the source's 
caustic crossings.  The binary lens produces a resonant caustic that is tilted away from the 
direction of the companion.  The source trajectory passes through the lower portion of the 
caustic, giving rise to the caustic-crossing features seen in the light curve.

% Table 4 ------------------------------------------------------
\begin{deluxetable*}{lllllll}
%\tabletypesize{\scriptsize}
\tablewidth{0pt}
\tablecaption{Lensing parameters of KMT-2024-BLG-1876, and KMT-2024-BLG-2379. \label{table:four}}
\tablehead{
\multicolumn{1}{c}{Parameter}           &
\multicolumn{1}{c}{KMT-2024-BLG-1876 }  &
\multicolumn{2}{c}{KMT-2024-BLG-2379 }  \\
\multicolumn{1}{c}{      }              &
\multicolumn{1}{c}{      }              &
\multicolumn{1}{c}{Close }              &
\multicolumn{1}{c}{Wide  }   
}
\startdata
 $\chi^2$               & $9522.0           $   &   $3357.7             $    &  $3341.0             $ \\
 $t_0$ (HJD$^\prime$)   & $522.047 \pm 0.020$   &   $553.178 \pm 0.010  $    &  $553.133 \pm 0.008  $ \\
 $u_0$                  & $0.1127 \pm 0.0047$   &   $0.01859 \pm 0.00051$    &  $0.01581 \pm 0.00060$ \\
 $\te$ (days)           & $37.88 \pm 1.04   $   &   $26.19 \pm 0.20     $    &  $25.10 \pm 1.04     $ \\
 $s$                    & $1.822 \pm 0.028  $   &   $0.524 \pm 0.016    $    &  $2.573 \pm 0.091    $ \\
 $q$                    & $0.0729 \pm 0.0041$   &   $0.0820 \pm 0.0069  $    &  $0.1561 \pm 0.0206  $ \\
 $\alpha$ (rad)         & $4.1439 \pm 0.0059$   &   $0.205 \pm 0.011    $    &  $0.143 \pm 0.007    $ \\
 $\rho$ ($10^{-3}$)     & $1.204 \pm 0.044  $   &   $2.041 \pm 0.039    $    &  $1.983 \pm 0.077    $ \\
\enddata
\tablecomments{
${\rm HJD}^\prime\equiv {\rm HJD}-2460000$.
}
\end{deluxetable*}
% --------------------------------------------------------------

\subsection{KMT-2024-BLG-1876 \label{sec:three-six}}

The lensing event KMT-2024-BLG-1876 was observed by the three major microlensing surveys. 
It was independently discovered by KMTNet and OGLE on 2024 July 19 (${\rm HJD}^\prime \sim 
510$), and an alert was issued by the MOA survey on 2024 August 1 (${\rm HJD}^\prime \sim 
523$).  The source, with a baseline magnitude of $I_{\rm base}=18.90$, lay in the overlap 
region of the KMTNet prime fields BLG02 and BLG42, which was monitored at a combined cadence 
of 15 minutes.

% Figure 6 ------------------------------------------------------
\begin{figure}[t]
\includegraphics[width=\columnwidth]{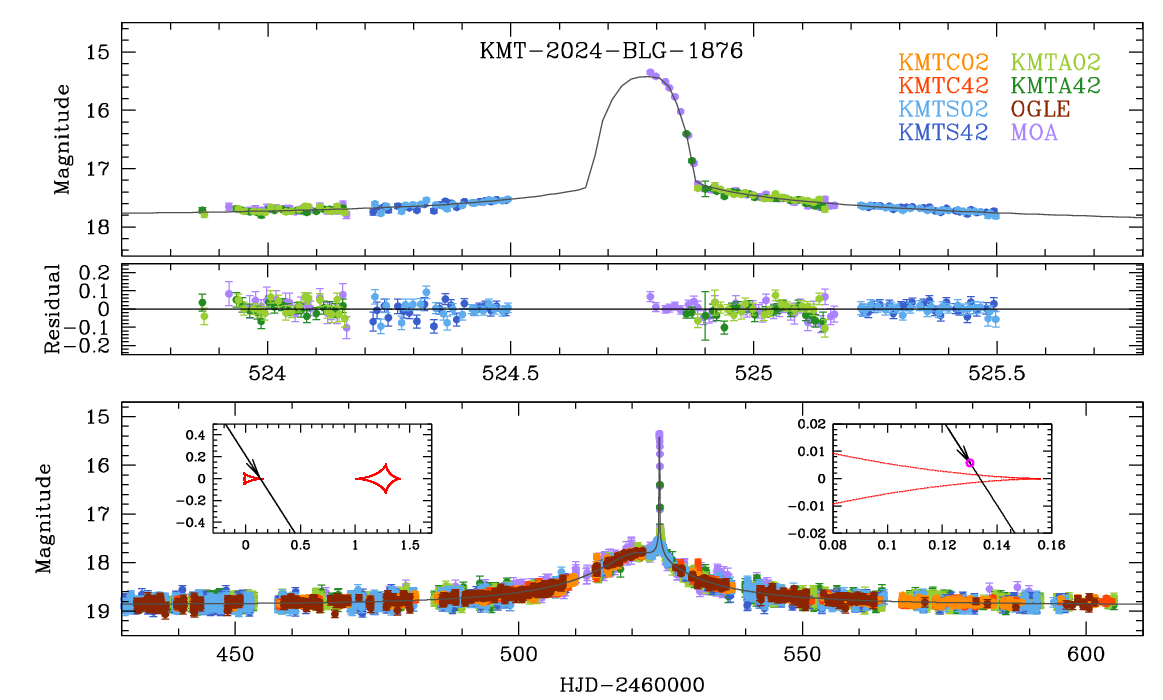}
\centering
\caption{
Lensing light curve of KMT-2024-BLG-1876. The layout of the insets in the bottom panel is 
the same as in Fig.~\ref{fig:five}.
}
\label{fig:six}
\end{figure}
% --------------------------------------------------------------

The lensing light curve is shown in Figure~\ref{fig:six}. It features a prominent bump lasting 
about half a day, centered at ${\rm HJD}^\prime \sim 525.8$. The declining side of this feature 
was covered by the combined MOA and KMTA data. Given the rapid fall-off, the bump is plausibly 
explained by the source grazing the tip of a caustic.

Light-curve modeling yields a best-fit wide-binary solution with parameters $(s, q) \sim (1.8, 
0.07)$ and an event timescale of $\te \sim 38$~days.  The small mass ratio strongly suggests 
that the lower-mass lens component is likely a BD. The full set of best-fit lensing parameters 
is listed in Table~\ref{table:four}, and the corresponding model curve is overplotted on the 
data in Figure~\ref{fig:six}. The densely sampled anomaly enables a precise measurement of the 
normalized source radius, $\rho \sim 1.20 \times 10^{-3}$. We also identify a local solution 
in the close-binary regime with $s \sim 0.64$, but it provides a substantially poorer fit than 
the wide solution, with $\Delta\chi^2 = 107.2$.

The insets in the bottom panel of Figure~\ref{fig:six} illustrate the lens-system geometry. 
In the wide-binary configuration, two caustics are produced, each located near one of the lens 
components. The observed anomaly arises from the source crossing the caustic associated with 
the more massive component. The source trajectory grazes the tip of the sharp on-axis caustic 
cusp on the companion side.  Because the time-normalized separation between the caustic entrance 
and exit is smaller than the normalized source radius, the perturbation appears as a single bump 
rather than the more typical caustic-crossing morphology consisting of two spikes and a U-shaped 
trough between them.

% Figure 7 ------------------------------------------------------
\begin{figure}[t]
\includegraphics[width=\columnwidth]{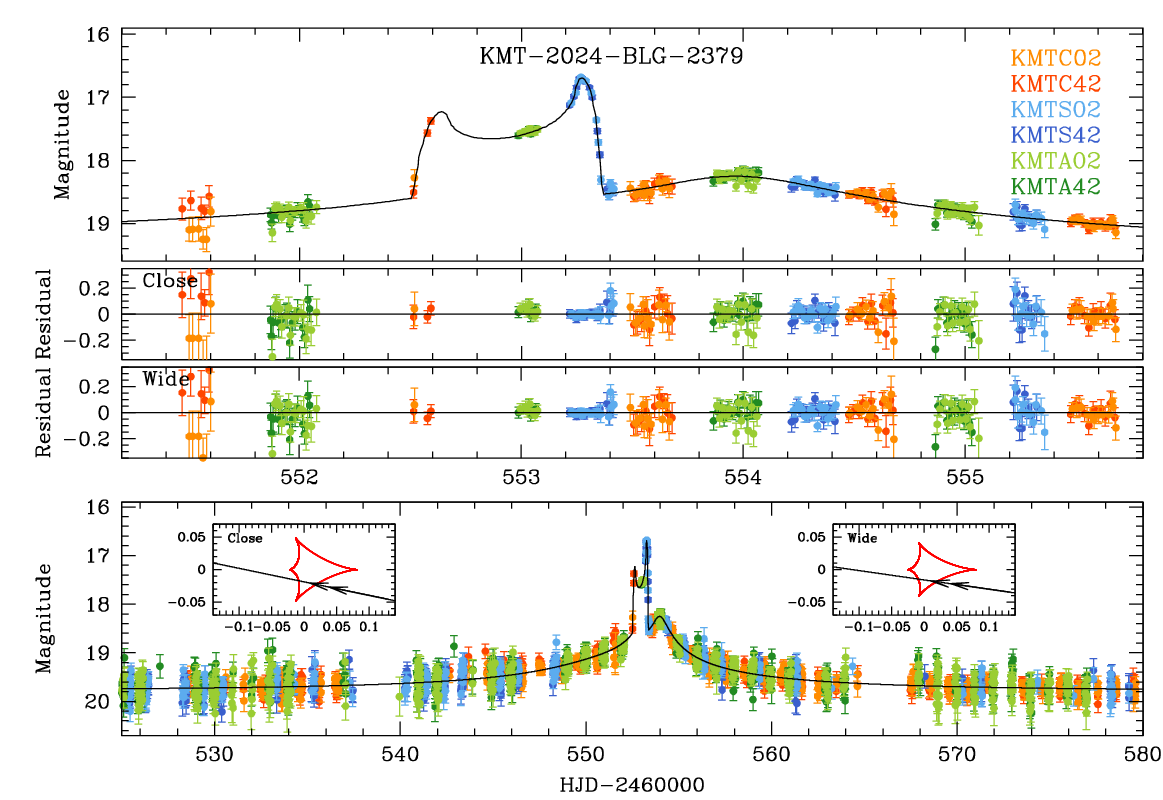}
\centering
\caption{
Lensing light curve of KMT-2024-BLG-2379. The two insets in the bottom panel 
show the configuration for the close and wide solutions.
}
\label{fig:seven}
\end{figure}
% --------------------------------------------------------------

\subsection{KMT-2024-BLG-2379\label{sec:three-seven}}

The lensing event KMT-2024-BLG-2379 was observed exclusively by the KMTNet survey. It was 
identified on 2024 August 29 (${\rm HJD}^\prime = 551$) during the rising phase of the 
magnification. Because the source lies in the overlap region of the KMTNet fields BLG02 
and BLG42, the light curve was densely sampled with a combined cadence of 15 minutes.

Figure~\ref{fig:seven} shows the light curve of the event. It exhibits a short-lived 
perturbation occurring about one day before the peak. The perturbation lasted for $\sim 1$~day 
and contains clear caustic-spike signatures.  The first spike is covered by the KMTC data, 
while the second is resolved by the KMTS data, with the intervening U-shaped trough covered 
by the KMTA data.

Modeling the light curve yields a pair of solutions arising from the close--wide 
degeneracy, which results from the similarity in the shapes of the central caustics 
produced by close ($s<1$) and wide ($s>1$) binaries \citep{Griest1998, Dominik1999, 
An2005, Chung2005}.  The binary-lens parameters are $(s, q)\sim (0.52, 0.08)$ for the 
close solution and $(s, q) \sim (2.57, 0.16)$ for the wide solution. The relatively 
small mass ratio suggests that the lens companion may be a BD.  The event timescale 
is similar for both solutions, $\te\sim 25$~days. Table~\ref{table:four} lists the full 
sets of lensing parameters for the two solutions.  The wide solution, whose mass ratio is 
$q_{\rm wide}/q_{\rm close}=1.9$ times that of the close solution, is favored over the 
close solution by $\Delta\chi^2=16.7$.  In Figure~\ref{fig:seven}, we overlay the model 
curve for the wide solution on the data.  The normalized source radius is tightly 
constrained to $\rho \sim 2.0\times 10^{-3}$.

The lens-system geometries for the two solutions are shown in the insets of the bottom panel 
of Figure~\ref{fig:seven}. The anomaly near the peak is produced by the source traversing the 
central caustic induced by the low-mass companion. After exiting the caustic, the source passes 
near the left on-axis cusp, producing a weak bump following the caustic exit.

% Table 5 ------------------------------------------------------
\begin{deluxetable*}{lllllll}
%\tabletypesize{\scriptsize}
\tablewidth{0pt}
\tablecaption{Lensing parameters of KMT-2025-BLG-0922, KMT-2025-BLG-1056, and  KMT-2025-BLG-2427.\label{table:five}}
\tablehead{
\multicolumn{1}{c}{Parameter}           &
\multicolumn{1}{c}{KMT-2025-BLG-0922 }  &
\multicolumn{1}{c}{KMT-2025-BLG-1056 }  &
\multicolumn{1}{c}{KMT-2025-BLG-2427 }  
}
\startdata
 $\chi^2$                   &  $2451.03          $   &  $3218.7             $  &    $585.4            $ \\
 $t_0$ (HJD$^\prime$)       &  $809.076 \pm 0.066$   &  $819.8212 \pm 0.0081$  &    $935.720 \pm 0.018$ \\
 $u_0$                      &  $  0.877 \pm 0.047$   &  $0.1506 \pm 0.0061  $  &    $0.1327 \pm 0.0025$ \\
 $\te$ (days)               &  $  2.680 \pm 0.081$   &  $4.58 \pm 0.16      $  &    $8.606 \pm 0.158  $ \\
 $s$                        &  $  1.966 \pm 0.013$   &  $2.233 \pm 0.028    $  &    $0.6746 \pm 0.0044$ \\
 $q$                        &  $  1.11 \pm 0.12  $   &  $1.58 \pm 0.12      $  &    $0.2416 \pm 0.0061$ \\
 $\alpha$ (rad)             &  $  4.492 \pm 0.025$   &  $5.1995 \pm 0.0058  $  &    $4.0015 \pm 0.0079$ \\
 $\rho$ ($10^{-3}$)         &  $  60.40 \pm 1.85 $   &  $42.51 \pm 1.94     $  &    $17.70 \pm 0.84   $ \\
 $ds/dt$ (yr$^{-1}$)        &     \nodata            &   \nodata               &    $1.16 \pm 0.44    $ \\
 $d\alpha/dt$ (yr$^{-1}$)   &     \nodata            &   \nodata               &    $2.63 \pm 0.30    $ \\
\enddata
\tablecomments{
${\rm HJD}^\prime\equiv {\rm HJD}-2460000$.
}
\end{deluxetable*}
% --------------------------------------------------------------

\subsection{KMT-2025-BLG-0922\label{sec:three-eight}}

The microlensing event KMT-2025-BLG-0922 was discovered by the KMTNet survey on 2025 May 2 
(${\rm HJD}^\prime=797$). The source has a baseline $I$-band magnitude of $I_{\rm base}=
17.83$, and the extinction toward the field is $A_I=1.26$. The event was independently 
identified by the PRIME survey on 2025 May 15 (${\rm HJD}^\prime=810$). Observations from 
both surveys were obtained at an approximately hourly cadence.

\begin{figure}[t]
\includegraphics[width=\columnwidth]{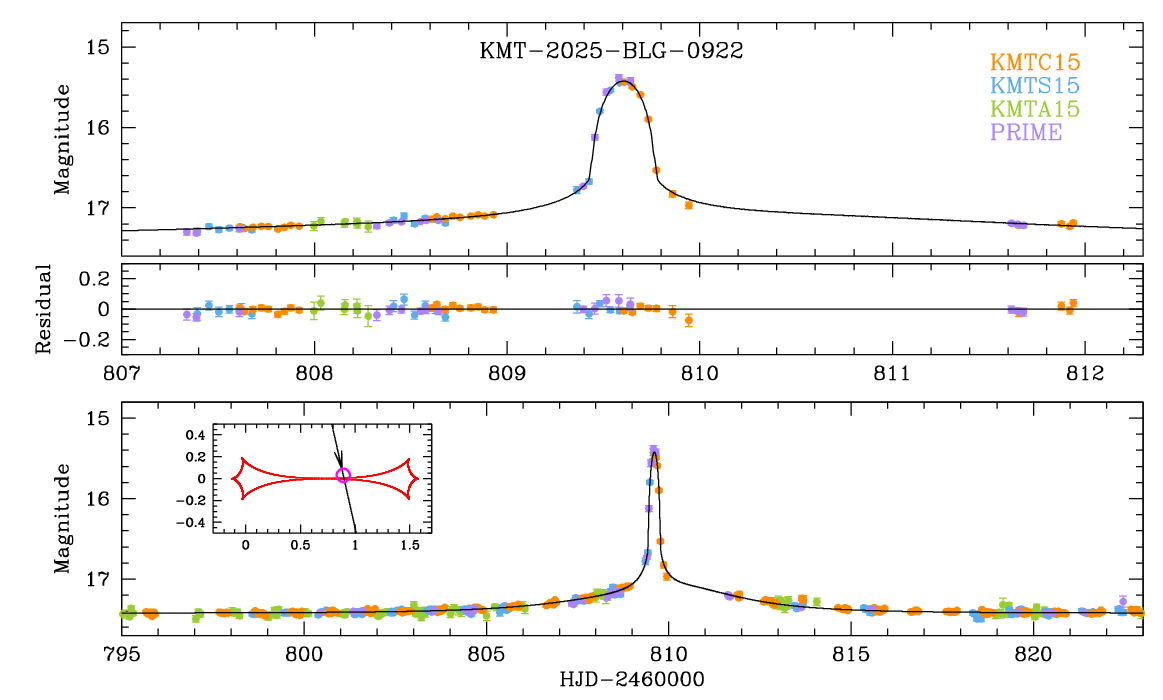}
\centering
\caption{
Light curve of KMT-2025-BLG-0922.
}
\label{fig:eight}
\end{figure}

Figure~\ref{fig:eight} shows the light curve of the event. It is characterized by a short 
timescale and a brief but pronounced anomaly centered at ${\rm HJD}^\prime = 809.6$. The 
anomaly is well resolved in the combined KMTNet and PRIME data. Given the high magnification 
relative to the baseline light curve, together with the rapid variation in magnification, 
the anomaly likely involves caustic crossings. However, the absence of a U-shaped trough 
between the rising and falling sides suggests that the source is larger than the separation 
between the caustic folds.

% Figure 8 ------------------------------------------------------
% --------------------------------------------------------------

Using a 2L1S model, we found a unique solution that reproduces both the overall light curve 
and the anomalous feature. The inferred binary parameters are $(s, q)\sim(1.97, 1.11)$. The 
full set of lensing parameters is listed in Table~\ref{table:five}, and the corresponding 
model light curve is shown in Figure~\ref{fig:eight}.  Although the masses of the lens 
components are comparable, we classify the lens as a brown-dwarf binary candidate because 
the event timescale is very short ($t_{\rm E} \sim 2.7$~days).  This interpretation is 
further supported by the large normalized source radius, $\rho\sim 60\times 10^{-3}$, which is 
substantially larger than typical values for events involving a giant source star. Such a 
large $\rho$ implies a small angular Einstein radius, and when combined with the short $\te$, 
suggests that both components of the lens are likely BDs.

The lens-system configuration is shown in the inset of the bottom panel of Figure~\ref{fig:eight}. 
Because the projected separation between the lens components is about twice the Einstein radius, 
the binary lens produces two caustics, each associated with one of the lens components. These 
caustics are connected by a thin bridge, which the source crossed with an incidence angle of 
$\sim77^\circ$. Owing to the narrowness of the bridge and the large source size, the resulting 
anomaly appears as a smooth bump, without sharp caustic spikes during the caustic crossings.

% Figure 9 ------------------------------------------------------
\begin{figure}[t]
\includegraphics[width=\columnwidth]{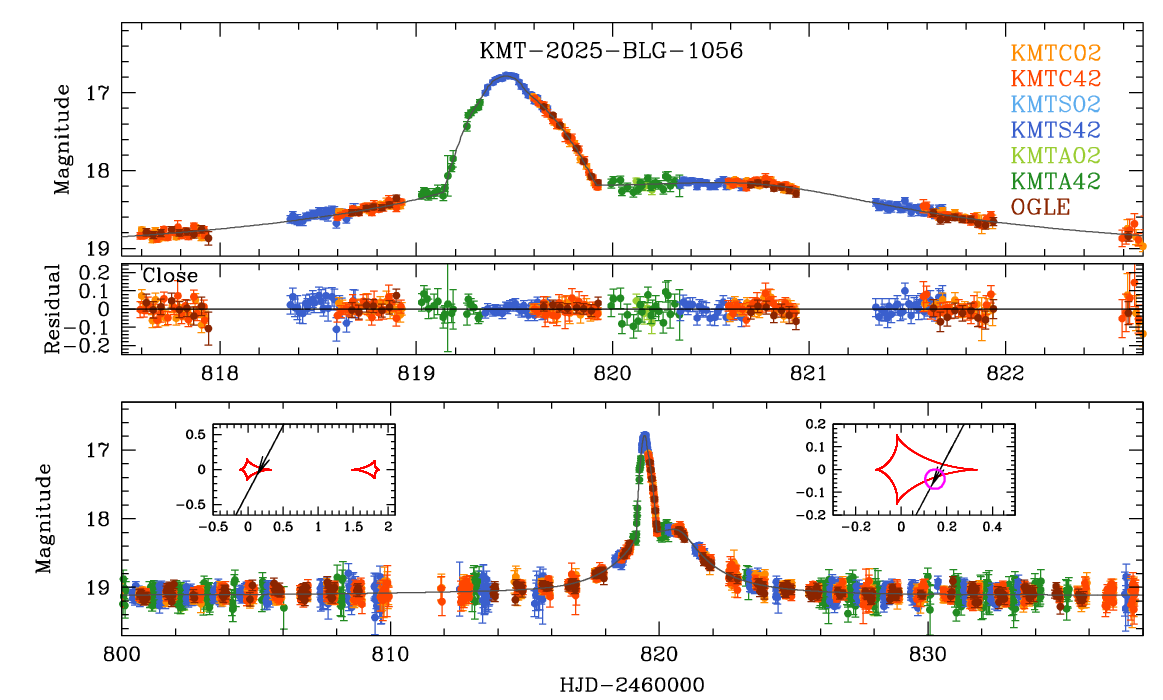}
\centering
\caption{
Light curve of KMT-2025-BLG-1056. The insets in the bottom panel adopt the same layout as 
in Fig.~\ref{fig:five}.
}
\label{fig:nine}
\end{figure}
% --------------------------------------------------------------

\subsection{KMT-2025-BLG-1056\label{sec:three-nine}}

The lensing event KMT-2025-BLG-1056 was observed by the KMTNet and OGLE surveys. It was 
first detected by KMTNet on 2025 May 23 (${\rm HJD}^\prime=820$) and independently 
identified by OGLE two days later. The source has a baseline magnitude of $I_{\rm base}=
18.91$. It lies in the overlap region of the KMTNet prime fields BLG02 and BLG42, which 
was monitored with a 15-minute cadence.  Because the source is located close to the Galactic 
center, at $(l, b) \sim (0^\circ\hskip-2pt .17, 1^\circ\hskip-2pt .14)$, the extinction toward 
the field is relatively high, $A_I=3.84$.

The light curve of KMT-2025-BLG-1056 is shown in Figure~\ref{fig:nine}. It is characterized 
by a prominent anomalous feature that occurs about one day before the peak. During the anomaly, 
the magnification rises and falls approximately linearly with time. This behavior differs from 
that of a typical caustic-crossing feature, which exhibits a pair of spikes separated by a 
U-shaped trough, and also from a cusp approach, which produces a smooth, rounded bump. Owing 
to the high-cadence observations, the anomaly is densely resolved by the combined data from 
OGLE and the three KMTNet telescopes.

Light-curve modeling yields a unique binary-lens solution with $(s, q)\sim (2.23, 1.58)$. 
Although the mass ratio is not small, we classify the lens as a binary BD candidate because 
the event timescale is very short, $\te\sim 4.6$~days. The full set of lensing parameters is 
provided in Table~\ref{table:five}. In addition, the large normalized source radius, $\rho\sim 
42.5\times 10^{-3}$, implies a small angular Einstein radius, further supporting the BD 
interpretation for the lens.  The model curve is plotted over the data in Figure~\ref{fig:nine}. 
In the modeling, we do not include the orbital motion of the lens because the model leaves 
little residual.

The lens-system configuration is shown in the inset of the bottom panel of Figure~\ref{fig:nine}. 
The wide binary lens produces two caustics, and the observed anomaly is generated when the source 
crosses the caustic located near the lower-mass component. The anomaly profile differs from a 
standard caustic-crossing pattern because the source traverses a protruding section of the 
caustic, where the source size exceeds the separation between the caustic folds.

\subsection{KMT-2025-BLG-2427\label{sec:three-eleven}}

The lensing event KMT-2025-BLG-2427 was first identified by the KMTNet survey on 2025 September 
15, corresponding to ${\rm HJD}^\prime = 933$. The event was also observed by the OGLE and MOA 
surveys. The source lies in the KMTNet field BKG15, which was monitored at a 1.0-hr cadence.

\begin{figure}[t]
\includegraphics[width=\columnwidth]{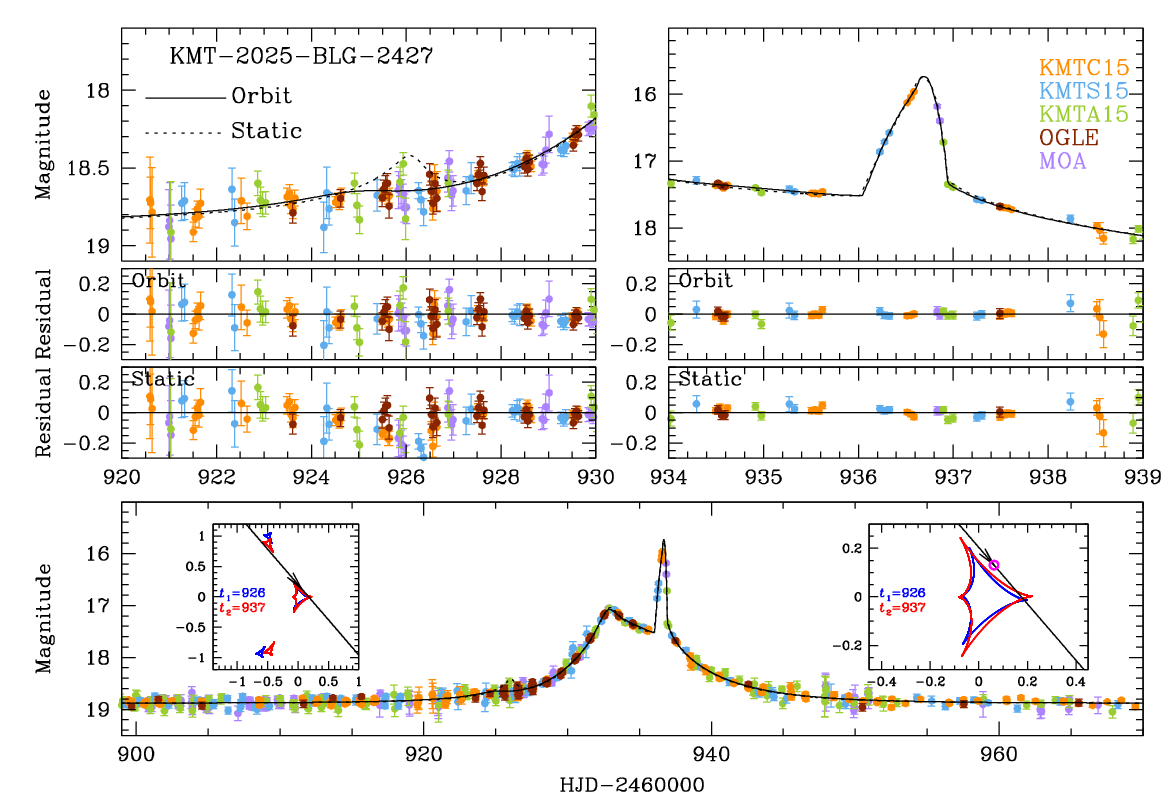}
\centering
\caption{
Light curve of KMT-2025-BLG-2427.  In the inset showing the lens-system 
configuration, the red and blue caustics correspond to two epochs ${\rm HJD}^\prime=926$ 
and 937.
}
\label{fig:ten}
\end{figure}

The light curve of the event is shown in Figure~\ref{fig:ten}. The peak region exhibits a complex 
anomaly pattern. First, the light curve displays a prominent bump centered at ${\rm HJD}^\prime 
\sim 936.7$.  The morphology of this major bump closely resembles that seen in KMT-2025-BLG-1056, 
suggesting that it was produced by the source passing over a cusp.  In addition, the light curve 
is noticeably asymmetric about the apparent peak at ${\rm HJD}^\prime \sim 933$.

Modeling of the light curve yields a unique 2L1S solution with binary parameters $(s, q) \sim 
(0.66, 0.27)$. We classify the lens as a BD companion because the event timescale is short, 
$\te \sim 8.1$~days. The model light curve under the standard interpretation (i.e., neglecting 
higher-order effects) is shown as the dotted curve in Figure~\ref{fig:ten}. In this model, 
the binary lens generates three caustics: a central caustic near the primary and two peripheral 
caustics. The major bump is produced when the source passes over the right-hand on-axis cusp of 
the central caustic.

For this event, lens orbital motion plays an important role. This is indicated by the fact that 
the standard model predicts the source trajectory to pass near the upper peripheral caustic, 
which should produce an anomaly around ${\rm HJD}^\prime \sim 926$, yet no such feature is 
observed. This discrepancy suggests that the source trajectory is effectively deflected by lens 
orbital motion, allowing it to avoid the caustic encounter. We therefore performed additional 
modeling that includes lens orbital motion. Given the short event timescale, we did not include 
the microlens-parallax effect in this analysis.

The higher-order (orbital) model (solid curve) and the residuals are shown in 
Figure~\ref{fig:ten}.  Including lens orbital motion substantially improves the fit, with 
$\Delta\chi^2 = 227.5$. For comparison, the residuals of the standard (non-orbital) model are 
also shown. The inferred orbital parameters are $(ds/dt, d\alpha/dt) \sim (1.16, 2.63)$~yr$^{-1}$, 
while the updated binary parameters, $(s, q) \sim (0.67, 0.24)$, and the event timescale, 
$\te \sim 8.6$~days, is similar to those of the standard model. The full lensing parameters 
of the orbital solution are listed in Table~\ref{table:five}.

% Figure 10 ------------------------------------------------------
% --------------------------------------------------------------

The lens-system configuration is illustrated in the two insets of the bottom panel of 
Figure~\ref{fig:ten}.  Because the caustic structure evolves under orbital motion, we 
plot the caustics at two epochs, ${\rm HJD}^\prime = 926$ and 937.  The geometry shows that 
the separation between the source trajectory and the upper peripheral caustic is sufficiently 
large that the peripheral caustic does not generate a detectable anomaly at the position 
predicted by the standard model.

% Figure 11 ------------------------------------------------------
\begin{figure*}[t]
\centering
\includegraphics[width=17.0cm]{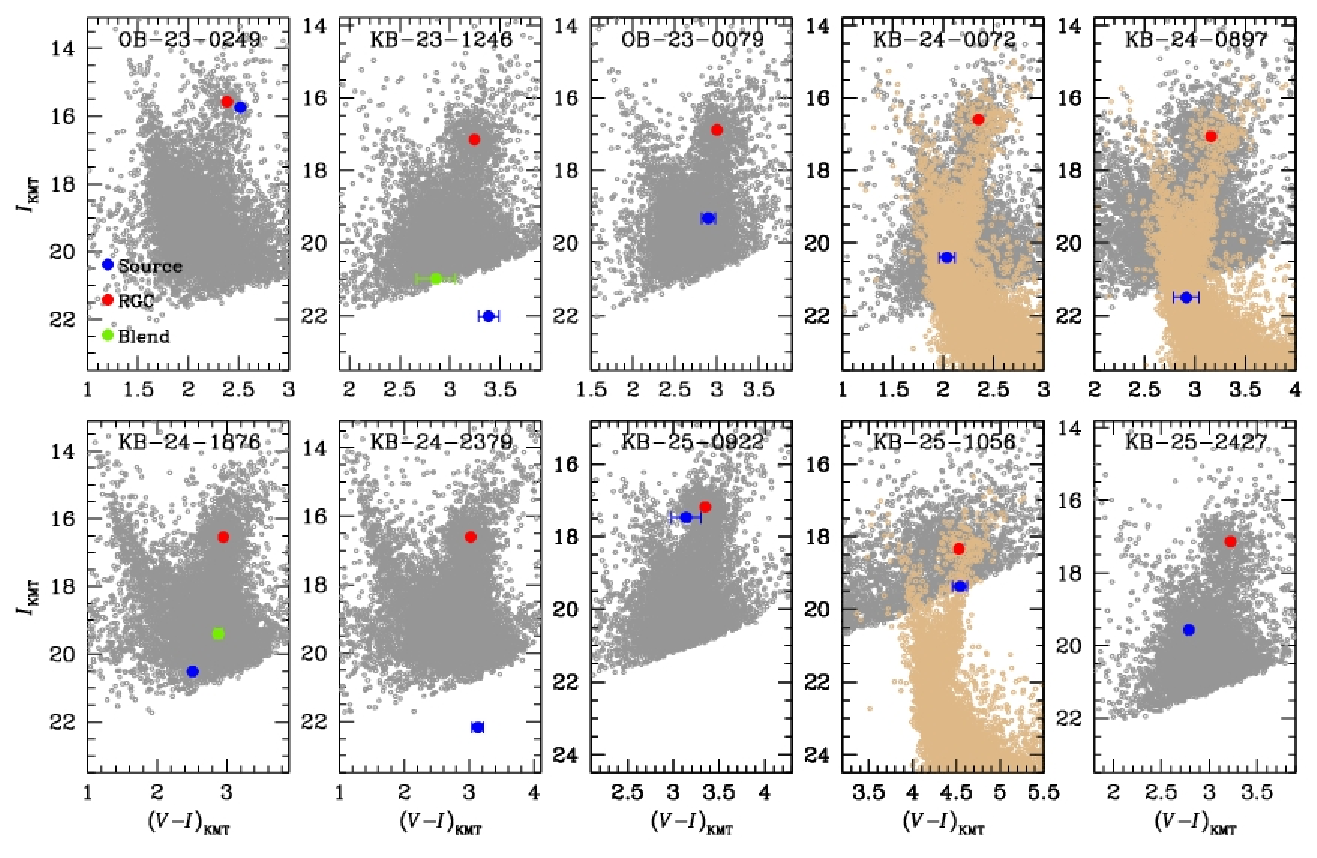}
\caption{
Locations of the source stars and the red giant clump (RGC) centroid in the instrumental 
color--magnitude diagrams. For events with measurable blended flux, the blend positions 
are also indicated.  For KMT-2024-BLG-0072, KMT-2024-BLG-0897, and KMT-2025-BLG-1056, the 
CMDs constructed by combining ground-based and HST observations.
}
\label{fig:eleven}
\end{figure*}
% --------------------------------------------------------------

% Table 6 ------------------------------------------------------
\begin{deluxetable*}{llllllllll}
%\tabletypesize{\scriptsize}
\tablewidth{0pt}
\tablecaption{Source parameters, angular Einstein radius, and relative lens-source proper motion.\label{table:six}}
\tablehead{
\multicolumn{1}{c}{Event}                    &
\multicolumn{1}{c}{$(V-I, I)_0$}             &
\multicolumn{1}{c}{Type}                     &
\multicolumn{1}{c}{$\theta_*$ ($\mu$as)}     &
\multicolumn{1}{c}{$\thetae$ (mas)}          & 
\multicolumn{1}{c}{$\mu_{\rm geo}$ (mas/yr)}         
}
\startdata
 OGLE-2023-BLG-0249  &   $(1.191\pm 0.040, 14.671\pm 0.020)$  &  K2III    &  $6.90\pm 0.56  $  &   $1.46\pm 0.12   $  &  $10.26 \pm 0.83 $  \\
 KMT-2023-BLG-1246   &   $(1.200\pm 0.107, 19.323\pm 0.020)$  &  K4.5V    &  $0.757\pm 0.097$  &   $0.437\pm 0.077 $  &  $7.27 \pm 1.28$    \\
 OGLE-2023-BLG-0079  &   $(0.957\pm 0.093, 16.791\pm 0.022)$  &  K2IV     &  $1.83\pm 0.21  $  &   \nodata            &  \nodata            \\
 KMT-2024-BLG-0072   &   $(0.745\pm 0.079, 18.142\pm 0.007)$  &  G6V      &  $0.770\pm 0.081$  &   $ > 0.06 $         &  $ > 3.0$           \\
 KMT-2024-BLG-0897   &   $(0.814\pm 0.128, 18.937\pm 0.019)$  &  G9V      &  $0.577\pm 0.084$  &   $0.152\pm 0.030 $  &  $5.18 \pm 1.04$    \\
 KMT-2024-BLG-1876   &   $(0.617\pm 0.052, 18.399\pm 0.020)$  &  F9V      &  $0.595\pm 0.052$  &   $0.505\pm 0.047$   &  $4.78 \pm 0.45$    \\
 KMT-2024-BLG-2379   &   $(1.170\pm 0.101, 19.999\pm 0.028)$  &  K4V      &  $0.530\pm 0.065$  &   $0.249\pm 0.032 $  &  $3.89 \pm 0.50$    \\
 KMT-2025-BLG-0922   &   $(0.850\pm 0.040, 14.770\pm 0.028)$  &  G3III    &  $4.12\pm 0.24  $  &   $0.068\pm 0.004 $  &  $9.29 \pm 0.56$    \\
 KMT-2025-BLG-1056   &   $(1.075\pm 0.084, 15.485\pm 0.098)$  &  K1III    &  $3.791\pm 0.415$  &   $0.089\pm 0.011 $  &  $7.11 \pm 0.88$    \\
 KMT-2025-BLG-2427   &   $(0.623\pm 0.044, 16.883\pm 0.020)$  &  F9V      &  $1.204\pm 0.100$  &   $0.068\pm 0.007 $  &  $2.89 \pm 0.28$    \\
\enddata
%\tablecomments{ }
\end{deluxetable*}
% --------------------------------------------------------------

% Table 7 ------------------------------------------------------
\begin{deluxetable*}{llccccccc}
%\tabletypesize{\scriptsize}
\tablewidth{0pt}
\tablecaption{Physical lens parameters.\label{table:seven}}
\tablehead{
\multicolumn{1}{c}{Event}                                 &
\multicolumn{1}{c}{Solution}                              &
\multicolumn{1}{c}{$M_1$ ($M_\odot$)}                     &
\multicolumn{1}{c}{$M_2$ ($M_\odot$)}                     &
\multicolumn{1}{c}{$\dl$ (kpc)}                           &
\multicolumn{1}{c}{$a_\perp$ (au)}                        & 
\multicolumn{1}{c}{$\boldsymbol{\mu}_{\rm L}$ (mas/yr)}   &
\multicolumn{1}{c}{$p_{\rm disk}$ }                       &
\multicolumn{1}{c}{$p_{\rm bulge}$}  
}
\startdata
 OGLE-2023-BLG-0249    &  $u_0>0$  &  $0.368 \pm 0.042        $    &  $0.0478 \pm 0.0056      $  &  $1.30 \pm 0.13       $   &  $3.23 \pm 0.32          $    &  $(-14.00, -6.56)$  & 100\%  &   0\%  \\  [0.8ex] 
                       &  $u_0<0$  &  $0.281 \pm 0.029        $    &  $0.0371 \pm 0.0038      $  &  $1.06 \pm 0.10       $   &  $2.61 \pm 0.24          $    &  $(-0.21, -5.59)$   &        &        \\  [0.8ex] 
 \hline                                                                                                                                                                                                          
 KMT-2023-BLG-1246     &  inner    &  $0.59^{+0.31}_{-0.30}   $    &  $0.009^{+0.005}_{-0.005}$  &  $6.66^{+1.01}_{-1.43}$   &  $2.57^{+0.39}_{-0.55}   $    &  \nodata            & 36\%   &  64\%  \\  [0.8ex] 
                       &  Outer    &  $0.59^{+0.31}_{-0.30}   $    &  $0.011^{+0.006}_{-0.006}$  &  $6.66^{+1.01}_{-1.43}$   &  $3.49^{+0.53}_{-0.75}   $    &  \nodata            &        &        \\  [0.8ex] 
 \hline                                                                                                                                                                                                         
 OGLE-2023-BLG-0079    &  Close    &  $0.57^{+0.37}_{-0.34}   $    &  $0.016^{+0.010}_{-0.010}$  &  $6.33^{+1.22}_{-1.82}$   &  $2.10^{+0.40}_{-0.60}   $    &  \nodata            & 43\%   &  57\%  \\  [0.8ex] 
                       &  Wide     &  $0.57^{+0.37}_{-0.34}   $    &  $0.021^{+0.014}_{-0.013}$  &  $6.33^{+1.22}_{-1.82}$   &  $3.18^{+0.61}_{-0.92}   $    &  \nodata            &        &        \\  [0.8ex] 
 \hline                                                                                                                                                                                                         
 KMT-2024-BLG-0072     &           &  $0.100^{+0.147}_{-0.057}$    &  $0.060^{+0.089}_{-0.035}$  &  $6.74^{+1.05}_{-1.15}$   &  $1.93^{+0.30}_{-0.33}   $    &  \nodata            & 24\%   &  76\%  \\  [0.8ex] 
 \hline                                                                                                                                                                                                         
 KMT-2024-BLG-0897     &           &  $0.15^{+0.23}_{-0.08}   $    &  $0.014^{+0.022}_{-0.008}$  &  $7.65^{+1.06}_{-1.13}$   &  $1.22^{+0.17}_{-0.18}   $    &  \nodata            & 23\%   &  77\%  \\  [0.8ex] 
 \hline                                                                                                                                                                                                         
 KMT-2024-BLG-1876     &           &  $0.70^{+0.30}_{-0.33}   $    &  $0.051^{+0.022}_{-0.024}$  &  $6.48^{+0.92}_{-1.42}$   &  $6.13^{+0.87}_{-1.35}   $    &  \nodata            & 45\%   &  55\%  \\  [0.8ex] 
 \hline                                                                                                                                                                                                          
 KMT-2024-BLG-2379     &  Close    &  $0.33^{+0.34}_{-0.18}   $    &  $0.027^{+0.027}_{-0.015}$  &  $7.49^{+0.96}_{-1.25}$   &  $1.02^{+0.13}_{-0.17}   $    &  \nodata            & 25\%   &  75\%  \\  [0.8ex] 
                       &  Wide     &  $0.33^{+0.34}_{-0.18}   $    &  $0.052^{+0.052}_{-0.029}$  &  $7.49^{+0.96}_{-1.25}$   &  $5.19^{+0.67}_{-0.87}   $    &  \nodata            &        &        \\  [0.8ex] 
 \hline                                                                                                                                                                                                         
 KMT-2025-BLG-0922     &           &  $0.047^{+0.090}_{-0.030}$    &  $0.042^{+0.081}_{-0.027}$  &  $7.44^{+1.01}_{-0.98}$   &  $1.00^{+0.14}_{-0.13}   $    &  \nodata            & 18\%   &  82\%  \\  [0.8ex] 
 \hline                                                                                                                                                                                                         
 KMT-2025-BLG-1056     &           &  $0.063^{+0.106}_{-0.036}$    &  $0.040^{+0.067}_{-0.023}$  &  $7.97^{+0.94}_{-0.92}$   &  $1.51^{+0.18}_{-0.18}   $    &  \nodata            & 1\%    &  99\%  \\  [0.8ex] 
 \hline                                                                                                                                                                                                         
 KMT-2025-BLG-2427     &           &  $0.09^{+0.083}_{-0.035}$     &  $0.014^{+0.020}_{-0.008}$  &  $8.03^{+1.17}_{-1.07}$   &  $0.432^{+0.063}_{-0.058}$    &  \nodata            & 10\%   &  90\%  \\  [0.8ex] 
\enddata
%\tablecomments{$\mu_{\rm helio}$ }
\end{deluxetable*}
% --------------------------------------------------------------

\section{Source star and Einstein radius\label{sec:four}}

In this section, we determine the source properties by deriving their de-reddened colors and 
magnitudes. This not only enables a characterization of the sources, but is also required to 
estimate the angular Einstein radius,
\begin{equation}
\theta_{\rm E} = {\theta_* \over \rho},
\label{eq4}
\end{equation}
\hskip-4pt
where the angular source radius $\theta_*$ is inferred from the source color and magnitude, 
and the normalized source radius $\rho$ is measured from light-curve modeling.

We obtain the de-reddened source color and magnitude, $(V-I, I)_0$, using a color--magnitude 
diagram (CMD) calibration relative to the red giant clump (RGC) in the same field 
\citep{Yoo2004}.  The RGC provides an excellent reference because its intrinsic color and 
absolute magnitude are well established for the Galactic bulge \citep{Bensby2013, Nataf2013}. 
Because the source and neighboring field stars lie behind nearly the same column of interstellar 
dust, this differential method yields a robust correction for extinction and reddening without 
requiring an absolute extinction map.

We first construct an instrumental CMD using KMTNet photometry of stars in the vicinity of 
the event. The source is then placed on this CMD using its instrumental color and magnitude. 
The unmagnified source fluxes in the $I$ and $V$ bands are determined by regressing the $I$- 
and $V$-band data against the model magnification, yielding the corresponding baseline 
(unmagnified) instrumental magnitudes.

We then identify the RGC centroid on the instrumental CMD. By comparing the observed RGC
centroid, $(V-I, I)_{\rm RGC}$, to its de-reddened value, $(V-I, I)_{{\rm RGC},0}$, we 
infer the reddening and extinction along the line of sight.  Assuming that the source 
suffers the same extinction as the RGC, the color excess and $I$-band extinction are
\begin{equation}
\eqalign{
 & E(V-I)  =  (V-I)_{\rm RGC} - (V-I)_{{\rm RGC},0}, \cr
 & A_I     =  I_{\rm RGC} - I_{{\rm RGC},0}.         \cr
\label{eq5}
}
\end{equation}
\hskip-9pt
The 
$I_{{\rm RGC},0}$ 
values are taken from Table~1 of \citet{Nataf2013}.
Applying these offsets to the source yields the de-reddened source color and magnitude,
$(V-I)_0$ and $I_0$.

The source colors of KMT-2024-BLG-0072, KMT-2024-BLG-0897, and KMT-2025-BLG-1056 could not 
be determined using the method described above, because the $V$-band observations either 
sparsely sample the light curve or suffer from large photometric uncertainties.  For these 
events, we instead combine the ground-based CMD with that constructed from Hubble Space 
Telescope (HST) observations \citep{Holtzman1998}, and estimate the source color by adopting 
the mean color of stars on either the main-sequence or the giant branch whose $I$-band offsets 
from the RGC lie within the uncertainty range of the source $I$-band magnitude.  For 
KMT-2025-BLG-0922, we confirmed that the source color and magnitude were consistent with 
that of the baseline object, and therefore (because the source is a giant), we adopted the 
baseline-object values for these quantities.

Figure~\ref{fig:eleven} shows the locations of the source stars relative to the RGC centroid 
on the instrumental CMD. For events in which the blend color and magnitude are measured, we 
also mark the positions of the blended light. Table~\ref{table:six} lists the de-reddened 
colors and magnitudes of the source stars for each event, along with their inferred spectral 
types.  We find that the sources in OGLE-2023-BLG-0249, KMT-2025-BLG-0922, and KMT-2025-BLG-1056 
are giants, while the source of OGLE-2023-BLG-0079 is a K-type subgiant.  The remaining 
events have main-sequence sources with spectral types spanning late F to mid K.

With $(V-I, I)_0$ in hand, we estimate the angular source radius.  Specifically, we derive 
$\theta_*$ from the surface-brightness relation of \citet{Kervella2004}.  Finally, we compute 
the angular Einstein radius using Eq.~(\ref{eq4}).  With the angular Einstein radius, the 
geocentric relative proper motion between the lens and source is computed by $\mu_{\rm geo} = 
\thetae/\te$.  The resulting values of $\theta_*$, $\theta_{\rm E}$, and $\mu_{\rm geo}$ for 
each event are presented in Table~\ref{table:six}.  For OGLE-2023-BLG-0079, we do not report 
$\thetae$ or $\mu_{\rm geo}$ because $\rho$ could not be measured. For KMT-2024-BLG-0072, for 
which only an upper limit on $\rho$ is constrained, we report lower limits on $\thetae$ and 
$\mu_{\rm geo}$.

\section{Lens mass and distance\label{sec:five}}

The primary physical lens parameters are the lens mass $M$ and distance $\dl$, which are 
related to the microlensing observables by
\begin{equation}
\te ={\thetae \over \mu};\qquad
\thetae = \sqrt{\kappa M \pi_{\rm rel}};\qquad
\pie = {\pi_{\rm rel} \over \thetae},
\label{eq6}
\end{equation}
\hskip-4pt
where $\kappa = 4G/(c^{2}{\rm au}) \simeq 8.144~{\rm mas}~M_\odot^{-1}$.  For OGLE-2023-BLG-0249, 
all three observables ($\te$, $\thetae$, and $\pie$) are measured.  In this case, the lens mass 
and distance can be determined directly using the relations of \citet{Gould2000},
\begin{equation}
M = {\thetae \over \kappa\pie };
\qquad 
\dl = {{\rm au} \over \pie\thetae + \pi_{\rm S}}.
\label{eq7}
\end{equation}

For events with incomplete observational constraints, we infer $M$ and $\dl$ using a Bayesian
analysis.  For this analysis, we adopt priors on the lens mass function $P(M)$, spatial 
density $P(\dl)$, and velocity distribution $P(\boldsymbol{v})$ based on the Galactic model 
of \citet{Jung2021} and the mass-function prescription of \citet{Jung2022}. We draw trial 
lenses with physical parameters $(M, \dl, \boldsymbol{v})$ from these priors and compute the 
corresponding microlensing observables $(t_{{\rm E},i}, \theta_{{\rm E},i})$. Each trial is 
then weighted by the likelihood,
\begin{equation}
\eqalign{
& {\cal L}\propto \exp\left( -{ \chi^{2} \over 2} \right);  \cr
& \chi^2 = \left( { t_{{\rm E},i}-t_{\rm E,obs} \over \sigma_{t_{\rm E}} } \right)^2 +
           \left( { \theta_{{\rm E},i}-\theta_{\rm E,obs} \over \sigma_{\theta_{\rm E}} } 
\right)^2, \cr
}
\label{eq8}
\end{equation}
\hskip-4pt
where $(t_{\rm E,obs}, \theta_{\rm E,obs})$ are the measured values of the observables, and
$(\sigma_{t_{\rm E}}, \sigma_{\theta_{\rm E}})$ are their associated uncertainties.  The 
posterior probability distribution is then obtained from the weighted ensemble as
\begin{equation}
P(M, \dl~\vert~t_{\rm E,obs}, \theta_{\rm E,obs})
\propto
\int {\cal L}\,P(M)\,P(\dl)\,P(\boldsymbol{v})\,d\boldsymbol{v},
\label{eq9}
\end{equation}
\hskip-4pt
from which we derive marginalized constraints on $M$ and $\dl$.

In Table~\ref{table:seven}, we summarize the estimated masses of the lens components ($M_1$ and 
$M_2$), the lens distance, and the projected separation ($a_\perp$) between the components. For 
all events, the median mass of the companion lies in the BD regime, $0.01 \lesssim M/M_\odot 
\lesssim 0.08$.  For the two cases of KMT-2025-BLG-0922 and KMT-2025-BLG-1056, the primary mass 
also falls within this range, indicating that the lenses are BD binary systems.  For the remaining 
events, the primary masses correspond to sub-solar main-sequence stars spanning $\sim 0.1~M_\odot$ 
to $\sim 0.70~M_\odot$.

For OGLE-2023-BLG-0249, we report the lens proper motion, $\boldsymbol{\mu}_{\rm L}$. 
We first compute the heliocentric relative lens--source proper motion,
\begin{equation}
\boldsymbol{\mu}_{\rm hel} = \boldsymbol{\mu}_{\rm geo} +
{\pi_{\rm rel} 
\over {\rm au}} \boldsymbol{v}_{\oplus,\perp},
\label{eq10}
\end{equation}
\hskip-3pt
where $\boldsymbol{v}_{\oplus,\perp}$ is the Earth’s projected velocity at the time of peak 
magnification. Using the source proper motion from the Gaia DR3 catalog \citep{Gaia2023},
$\boldsymbol{\mu}_{\rm S}=(\mu_N,\mu_E)=(-7.272\pm0.117, -0.550\pm0.159)$~mas/yr,
we then compute the lens proper motion as
\begin{equation}
\boldsymbol{\mu}_{\rm L} = \boldsymbol{\mu}_{\rm hel} + \boldsymbol{\mu}_{\rm S}.
\label{eq11}
\end{equation}
\hskip-6pt
The inferred lens proper motions for the $u_0>0$, $\boldsymbol{\mu}_{\rm L}\sim(-1.40, -6.6)$~mas/yr,
and $u_0<0$, $\boldsymbol{\mu}_{\rm L}\sim(-0.21, -5.6)$~mas/yr, solutions differ substantially, 
suggesting that the corresponding lenses may belong to different Galactic populations (e.g., thick 
disk versus thin disk). Although the thin-disk interpretation may be favored, high-resolution 
adaptive-optics imaging is required to discriminate between the two solutions.

Also listed in Table~\ref{table:seven} are the probabilities that the lens resides in the disk 
($p_{\rm disk}$) or the bulge ($p_{\rm bulge}$).  For OGLE-2023-BLG-0249, the lens is very likely 
to be in the disk. In contrast, the lenses of 
KMT-2023-BLG-1246,
KMT-2024-BLG-0072,
KMT-2024-BLG-0897,
KMT-2024-BLG-2379,
KMT-2025-BLG-0922,
KMT-2025-BLG-1056, and
KMT-2025-BLG-2427 are likely to be in the bulge with $p_{\rm bulge} > 60\%$.  For OGLE-2023-BLG-0079 
and KMT-2024-BLG-1876, $p_{\rm disk}$ and $p_{\rm bulge}$ are comparable.

% Figure 10 ------------------------------------------------------
\begin{figure}[t]
\includegraphics[width=\columnwidth]{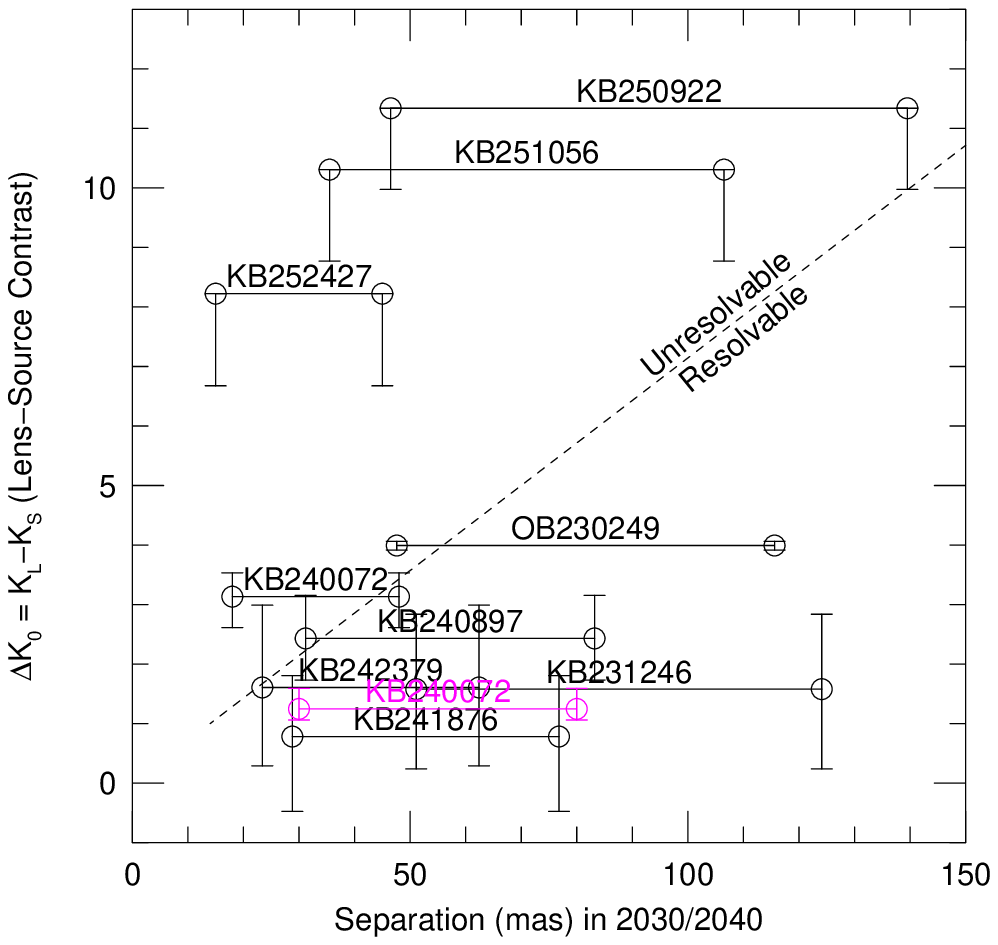}
\centering
\caption{
Feasibility of mass and distance measurements from future AO observations on the European 
Extremely Large Telescope.
}
\label{fig:twelve}
\end{figure}
% --------------------------------------------------------------

\section{Masses and Distances from AO Imaging} \label{sec:six}

High-resolution imaging, for example with the AO on the European Extremely Large Telescope (EELT), 
can plausibly yield mass and distance measurements of most of the primaries of the lens systems 
analyzed in this work, and thereby (after multiplying the primary mass by $q$), the masses of 
the BDs themselves. The method was first applied by \citet{Batista2015} and \citet{Bennett2015} 
and was explored more thoroughly by \citet{Gould2022}.

Figure~\ref{fig:twelve} illustrates the feasibility of applying this method to nine of the ten
systems listed in Tables~\ref{table:six} and \ref{table:seven}.  We exclude OGLE-2023-BLG-0079 
because neither has its proper motion been measured nor is an upper limit.  For each 
event we calculated the separation in 2030 (estimated first light of EELT) and also a decade 
later, in 2040.  We used the $(V-I,I)_0$ columns from Table~\ref{table:six} and the tables of 
\citet{Bessell1988} to estimate $K_{0,S}$ of the source. Next we used three values of the primary 
mass (median and $1\,\sigma$ limits) from Table~\ref{table:seven}, together with $\thetae$ from 
Table~\ref{table:six}, and the BD/primary mass ratio, $q$, to estimate the system distance for each 
adopted mass to find the corresponding lens distance, i.e., $\dl = {\rm au}/[\theta_{\rm E}^2/
\kappa M(1+q) + 0.125~{\rm mas}]$.  We combined this with the lens $K$-band absolute magnitude, 
derived from the mass-luminosity relation shown in Figure~22 of \citet{Benedict2016}, to find 
$K_{0,L} =M_K + 5\log (D_{\rm L}/10~{\rm pc})$. And finally, we plotted the $K$-band contrast 
$\Delta K =K_{0,L}-K_{0,S}$ for each of the three masses.

There were two exceptions to this general procedure. First, for OGLE-2023-BLG-0249, we used 
the heliocentric proper motion (rather than geocentric) because this governs the actual 
multi-year separation. For the other events, we used the geocentric proper motion as the best 
available proxy. Second, for KMT-2024-BLG-0072, we have only lower limits, $\thetae > 0.06$~mas 
and $\mu_{\rm geo} > 3.0$~mas/yr. Hence, we used these limits as our adopted values in our 
main calculation (shown in black in Figure~\ref{fig:twelve}). However, we also show separately 
(in magenta) a larger pair of plausible values, $\thetae=0.10$~mas and $\mu_{\rm geo}=5.0$~mas/yr.  
Finally, we only show the primary brightness in the stellar-mass range. If the primary itself is 
a BD, this method will only give an upper limit on its mass (and so the mass of its BD companion).

Because the instrument (MICADO) is not yet built, we do not yet know its performance 
specifications. In lieu of this, we show as a dashed line a rough estimate of the lower 
limit on source--lens separations that can be resolved.

Figure~\ref{fig:twelve} shows that six of these nine systems will likely be accessible to mass
measurement by 2040 and most likely well before that. Three of the systems KMT-2025-BLG-0922, 
KMT-2025-BLG-1056, and KMT-2025-BLG-2427, will likely remain inaccessible to this technique 
for several decades.

\section{Summary and Conclusions} \label{sec:seven}

We present detailed light-curve analyses of ten binary-lens microlensing events observed 
during the 2023--2025 seasons, selected as candidate BD companions in binary lens systems. 
The event sample comprises 
OGLE-2023-BLG-0249, 
KMT-2023-BLG-1246, 
OGLE-2023-BLG-0079, 
KMT-2024-BLG-0072, 
KMT-2024-BLG-0897, 
KMT-2024-BLG-1876, 
KMT-2024-BLG-2379, 
KMT-2025-BLG-0922,
KMT-2025-BLG-1056, and 
KMT-2025-BLG-2427.

For each event, we derived binary-lens parameters using 2L1S modeling and examined relevant 
degeneracies where applicable. For events exhibiting resolved caustic-related features in 
their light curves, we measured the angular Einstein radius by combining the normalized 
source radius with the angular source radius, which was inferred from the de-reddened source 
color and magnitude. For one event, OGLE-2023-BLG-0249, we additionally measured the microlens 
parallax.

We then inferred the physical properties of the lens systems. For OGLE-2023-BLG-0249, simultaneous 
measurements of the three lensing observables $(\te, \thetae, \pie)$ enabled a direct determination 
of the lens masses and distance. For the remaining events, for which the lensing observables were 
only partially measured, we performed a Bayesian analysis with Galactic priors to obtain posterior 
distributions for the component masses and the lens distance.

Our main results are as follows.
\begin{enumerate}
\item \textbf{BD companions:} 
For all events, the inferred companion lens masses have medians in the BD regime, $0.01 
\lesssim M_2/M_\odot \lesssim 0.08$.  This confirms that binary-lens events with small mass 
ratios (and/or small $\theta_{\rm E}$ when measurable) provide an efficient channel to identify 
BD companions in microlensing surveys.
\item \textbf{Binary BD lenses:} 
KMT-2025-BLG-0922 and KMT-2025-BLG-1056 have primary lens masses that are also consistent with 
the BD regime, indicating that the lenses are likely binaries 
composed of two BDs.
\item \textbf{Lens distances and separations:} 
The inferred lens distances span a wide range, from $\sim 1$ to 8~kpc, while the projected 
separations extend from sub-au to multi-au scales, depending on the event and the underlying 
degeneracy class. These systems therefore probe BD companions across a broad range of Galactic 
environments and binary configurations.
\item \textbf{Degeneracies and higher-order effects:} 
Several events exhibit well-known binary-lens degeneracies (close--wide; inner--outer), and at 
least one case (KMT-2025-BLG-2427) requires lens orbital motion to reproduce the observed 
light-curve structure. These features highlight the importance of dense temporal coverage and 
higher-order modeling for robust physical inference in short events.
\end{enumerate}

Overall, this sample demonstrates that current high-cadence survey data can deliver a growing and
well-characterized population of microlensing BD companions, including BD--BD binaries, extending
BD demographics to faint and distant systems inaccessible to flux-limited techniques. Future
high-resolution follow-up imaging (to measure lens flux and/or relative proper motion) and
continued survey coverage will be particularly valuable for tightening mass constraints in events
lacking parallax or $\rho$ measurements, and for refining the statistical properties (mass ratios
and separations) of microlensing-selected BD companions.

An additional feature of our results is that 9 out of the 10 events are found to have
lens systems preferentially located in the Galactic bulge. This apparent concentration
does not necessarily imply an intrinsic overabundance of BD companions in the bulge,
but instead reflects a combination of observational and methodological effects. First,
microlensing surveys toward the Galactic bulge predominantly monitor bulge source stars,
which increases the probability of detecting events involving bulge lenses. Second, the
microlensing optical depth is highest along these lines of sight, naturally favoring
lens--source configurations in which both components reside in the bulge. Third, our
selection criteria favor events with measurable finite-source effects, which tend to
arise in systems with relatively small angular Einstein radii. For a given source distance,
this condition is more readily satisfied when the lens is located close to the source,
i.e., in the bulge. Finally, the Bayesian analysis incorporates Galactic priors that
account for the density and kinematic distributions of disk and bulge populations, and
the combination of these priors with the observed $(t_{\rm E}, \theta_{\rm E})$ values
often leads to posterior distributions that favor bulge lenses. Therefore, the predominance
of bulge lenses in our sample should be understood as a consequence of these combined
effects rather than as evidence for a population-level trend.

% --------------------------------------------------------------
\begin{acknowledgements}
% Han C. 
C.H. was supported by the Korea Astronomy and Space Science Institute under the R\&D program 
(Project No. 2026190401) supervised by the Ministry of Science and ICT.
% LeeCU --
Work by C.U.Lee research was supported by the Korea Astronomy and Space Science Institute under 
the R\&D program (Project No. 2025-1-830-05) supervised by the Ministry of Science and ICT.
% KMTNet
This research has made use of the KMTNet system operated by the Korea Astronomy and Space Science 
Institute (KASI) at three host sites of CTIO in Chile, SAAO in South Africa, and SSO in Australia. 
Data transfer from the host site to KASI was supported by the Korea Research Environment Open 
NETwork (KREONET). 
% KASI support for Han 
% OGLE 
The OGLE project has received funding from the Polish National Science
Centre grant OPUS-28 2024/55/B/ST9/00447 awarded to AU.
% MOA -----
%The MOA project is supported by JSPS KAKENHI Grant Number JP16H06287, JP22H00153 and 23KK0060.
% Clement Ranc
%C.R. was supported by the Research fellowship of the Alexander von Humboldt Foundation.
%Chinese collaborator 
H.Y. and W.Z. acknowledge support by the National Natural Science Foundation of China (Grant 
No. 12133005). 
H.Y. acknowledge support by the China Postdoctoral Science Foundation (No. 2024M762938).  
\end{acknowledgements}

%%\appendix
%%Appendices can be broken into separate sections just like in the main text.
%%The only difference is that each appendix section is indexed by a letter
%%(A, B, C, etc.) instead of a number.  Likewise numbered equations have
%%the section letter appended.  Here is an equation as an example.
%%\begin{equation}
%%I = \frac{1}{1 + d_{1}^{P (1 + d_{2} )}}
%%\end{equation}
%%Appendix tables and figures should not be numbered like equations. Instead
%%they should continue the sequence from the main article body.

\bibliography{references}
\bibliographystyle{aasjournalv7}

\end{document}